\documentclass{jfm}
\usepackage{graphicx}
\usepackage{natbib}
\usepackage{amssymb}
\usepackage{amsbsy}
\usepackage{amsmath}
\usepackage{graphics}
\usepackage{subfigure}
\usepackage{dcolumn}
\usepackage{bm}
\usepackage{hyperref}
\usepackage{xcolor}
\usepackage{soul}
\usepackage{color}

\newcommand{\bc}{\begin{center}}
\newcommand{\ec}{\end{center}}

\newcommand{\be}{\begin{equation}}
\newcommand{\ee}{\end{equation}}
\newcommand{\bea}{\begin{eqnarray}}
\newcommand{\eea}{\end{eqnarray}}

\title[]{Is Chaotic Advection Inherent to Heterogeneous Darcy Flow?}

\author[Daniel R. Lester, Michael G. Trefry, Guy Metcalfe, Marco Dentz]%
{Daniel R. Lester$^1$%
  \thanks{Email address for correspondence: daniel.lester@rmit.edu.au},\ns
Michael G. Trefry$^2$\break
Guy Metcalfe$^3$,\ns Marco Dentz$^4$}

\affiliation{$^1$School of Engineering, RMIT University, Melbourne, Australia\\[\affilskip]
$^2$Independent Researcher, Perth, Australia\\[\affilskip]
$^3$Swinburne University of Technology, Melbourne, Australia\\[\affilskip]
$^4$ Spanish National Research Council (IDAEA-CSIC), 08034 Barcelona, Spain\\[\affilskip]
}

\pubyear{2024}
\volume{}
\pagerange{}
\date{?; revised ?; accepted ?. - To be entered by editorial office}
\begin{document}

\maketitle

\begin{abstract}
At all scales, porous materials stir interstitial fluids as they are advected, leading to complex distributions of matter and energy. Of particular interest is whether porous media naturally induce chaotic advection at the Darcy scale, as these stirring kinematics profoundly impact basic processes such as solute transport and mixing, colloid transport and deposition and chemical, geochemical and biological reactivity. While many studies report complex transport phenomena characteristic of chaotic advection in heterogeneous Darcy flow, it has also been shown that chaotic dynamics are prohibited in a large class of Darcy flows. In this study we rigorously establish that chaotic advection is inherent to steady 3D Darcy flow in all realistic models of heterogeneous porous media. Anisotropic and heterogenous 3D hydraulic conductivity fields generate non-trivial braiding of streamlines, leading to both chaotic advection and (purely advective) transverse dispersion. We establish that steady 3D Darcy flow has the same topology as unsteady 2D flow, and so use braid theory to establish a quantitative link between transverse dispersivity and Lyapunov exponent in heterogeneous Darcy flow. We show that chaotic advection and transverse dispersion occur in both anisotropic weakly heterogeneous and in heterogeneous weakly anisotropic conductivity fields, and that the quantitative link between these phenomena persists across a broad range of conductivity anisotropy and heterogeneity. The ubiquity of macroscopic chaotic advection has profound implications for the myriad of processes hosted in heterogeneous porous media and calls for a re-evaluation of transport and reaction methods in these systems.
\end{abstract}

\begin{keywords}
Darcy flow, porous media, chaotic advection, topological braiding
\end{keywords}

\section{Introduction}\label{sec:intro}
Porous media abound in both nature and engineered systems. From biological tissues to geological media and engineered materials, these media encompass a vast range of applications~\citep{Bear:1972aa,Cushman:2013aa}. Porous materials play host to a range of fluid-borne phenomena, including mixing~\citep{Villermaux:2012aa,Dentz:2023aa}, dispersion~\citep{saffman1959theory,Bear:1972aa,Gelhar:1983aa,Dagan:1989aa}, transport~\citep{brenner:book} and reaction~\citep{Anna:2013ab,Rolle:2019aa,Valocchi:2019aa} of colloids, chemical and biological species \citep{alonso2019transport,dentz2022dispersion}. These phenomena are governed by the Lagrangian kinematics of the interstitial fluid, which provide an \emph{advective template} that organises the spatial structures upon which they play out. At all scales, porous materials act to stir fluids as they are advected through~\citep{Villermaux:2012aa,Dentz:2023aa}, leading to highly striated material distributions which can have profound impacts upon these fluid-borne phenomena.

For example, solute mixing and transport can be either significantly accelerated or retarded by the Lagrangian kinematics of the flow, leading to the existence of e.g. isolated fast or slow mixing regions and transport ``barriers'' or ``highways''~\citep{Haller:2015aa,Wu:2024aa}. Similarly, the kinetics of reactive solutes are governed by transport and mixing, including the formation of reaction ``hot spots'' due to localised fluid deformation~\citep{Rolle:2019aa,Valocchi:2019aa}. It is impossible to understand, quantify and predict these processes without resolving their underlying Lagrangian kinematics~\citep{Metcalfe:2023aa}.

One important class of such kinematics is \emph{chaotic advection}~\citep{Arnold:1965aa,Aref:1984aa} where fluid stirring motions (stretching and folding) yield highly striated distributions of matter and energy that can fundamentally alter these fluid-borne phenomena~\citep{Aref:2017aa}. For example, solute mixing is singular under chaotic advection~\citep{Cerbelli:2017aa,Fereday:2004aa}  in that scalar dissipation persists in the limit of vanishingly small molecular diffusivity. Similarly, chaotic advection qualitatively augments both longitudinal~\citep{Jones:1994aa} and transverse dispersion~\citep{Lester:2014ab}. Colloidal transport is also strongly augmented by chaotic advection, leading to formation of particle traps, repellers and deposition hot spots on fluid boundaries~\citep{Ouellette:2008aa,Haller:2008aa}. Chaotic advection also fundamentally alters chemical reactions~\citep{Tel:2005aa,Neufeld:2009aa} and biological activity~\citep{Tel:2000aa,Karolyi:2000aa}, leading , e.g., singularly enhanced kinetics, coexistence of competitive species and altered stability characteristics. Hence, detection and quantification of chaotic advection in porous media flows is key to understanding, prediction and upscaling the myriad fluidic processes hosted in porous media.

Chaotic advection is inherent to steady three-dimensional (3D) pore-scale flow. It is now firmly established  ~\citep{Lester:2013aa,Lester:2016ab,Turuban:2018aa,Turuban:2019aa,Kree:2017aa,Souzy:2020aa,Heyman:2021aa} that pore-scale chaotic advection arises in almost all porous media, ranging from granular matter to open pore networks. At the Darcy scale, chaotic advection has been predicted or observed in both natural~\citep{Trefry:2019aa,Wu:2020aa,Metcalfe:2023aa,Wu:2024aa} and engineered~\citep{Cho:2019aa,Metcalfe:2010aa,Mays:2012aa} transient Darcy flows. However chaotic advection has not been explicitly detected under steady 3D Darcy flow, despite consistent observations of complex Lagrangian kinematics~\citep{Chiogna:2014aa,Ye:2015aa}. Conversely, several recent studies~\citep{Lester:2021aa, Lester:2022aa} have firmly established that chaotic dynamics cannot occur under steady 3D Darcy flow with smooth ($C^1$-continuous) isotropic hydraulic conductivity fields. As such, the prevalence and nature of chaotic advection in steady heterogeneous Darcy flow is not understood.

In this study we rigorously establish that chaotic advection is inherent to all steady Darcy flows that are generated by faithful representations of heterogeneous porous media. This is important as most porous media are inherently heterogeneous at the Darcy scale~\citep{Bear:1972aa,Cushman:2013aa}, and solute transport at this scale is typically advection dominated~\citep{Delgado:2007aa,Bear:1972aa}, hence chaotic advection profoundly impacts solute transport, mixing and reactions. We consider the simplest non-trivial system of steady 3D Darcy flow arising from a mean potential gradient in unbounded systems with smooth, finite hydraulic conductivity fields. Although this precludes systems with stagnation points, non-smooth conductivity fields or impermeable inclusions, our focus here is to understand basic Darcy flow. We establish that chaotic advection arises in the simplest physically plausible heterogeneous conductivity structures, even for weakly heterogeneous porous materials. Conversely strongly heterogeneous porous media exhibits a Lyapunov exponent close to the theoretical upper bound for steady 3D flow. We elucidate the underlying mechanisms and establish a quantitative link between the strength of chaotic mixing and (purely advective) transverse dispersion. Application of this link to experimental dispersion data also indicates chaotic mixing is significant and ubiquitous in heterogeneous Darcy flow.

This work is organized as follows. In \S\ref{sec:background} we show that all realistic models of heterogeneous media must have anisotropic hydraulic conductivity fields. The braiding of streamlines in Darcy flow and the connection with chaotic advection are then considered in \S\ref{sec:braiding}, including development of a quantitative link between transverse dispersion and Lyapunov exponent in random braiding flows. Numerical simulations are performed in \S\ref{sec:numerics}, including anisotropic Darcy flow with variable heterogeneity and heterogeneous Darcy with varying anisotropy. In \S\ref{sec:mechanisms_implications} the mechanisms generating chaotic advection in Darcy flow and implications for solute mixing are discussed. Conclusions are given in \S\ref{sec:conclusions}. 


\section{Kinematics of Porous Media Flows}\label{sec:background}

\subsection{Kinematics of Pore and Darcy Scale Flow}

We first consider the kinematics of steady 3D pore-scale flow before upscaling to Darcy flow. Steady 3D pore scale flow is described by the Stokes equations
\begin{equation}
\mu\nabla^2\hat{\mathbf{v}}(\mathbf{x})-\nabla\hat{p},\quad \nabla\cdot\hat{\mathbf{v}}=0,\,\,\,\,\mathbf{x}\in\Omega_f,\label{eqn:Stokes}
\end{equation}
subject to no-slip conditions $\hat{\mathbf{v}}|_{\partial\Omega_{fs}}=\mathbf{0}$ at the pore boundary $\partial\Omega_{fs}$. Here $\hat{\mathbf{v}}$ is the pore-scale velocity, $\hat{p}$ the pore-scale pressure field, $\Omega_f$, $\Omega_s$ respectively are the fluid (pore) and solid (grain) domains. For steady 3D flow, the topological complexity of the boundary $\partial\Omega_{fs}$ generates pore-scale chaotic advection~\citep{Lester:2013aa}. Upscaling by suitably averaging of (\ref{eqn:Stokes}) (denoted by $\langle\cdot\rangle_p$) yields the Darcy equation~\citep{Bear:1972aa}
\begin{equation}
\mathbf{v}(\mathbf{x})=-\mathbf{K}(\mathbf{x})\cdot\nabla \phi(\mathbf{x}),\quad \nabla\cdot\mathbf{v}=0,\,\,\,\,\mathbf{x}\in\Omega,\label{eqn:Darcy}
\end{equation}
where $\Omega\equiv\langle\Omega_f\cup\Omega_s\rangle_p$ denotes the Darcy-scale porous medium, $\mathbf{v}\equiv\langle\hat{\mathbf{v}}\rangle_p$ is the (upscaled) Darcy-scale velocity, $\phi\equiv\langle\hat{p}\rangle_p$ the upscaled potential field, and $\mathbf{K}(\mathbf{x})$ is the hydraulic conductivity tensor field that captures viscous drag due to no-slip conditions on $\partial\Omega_{fs}$. 

As this pore boundary generates chaotic advection, omission of $\partial\Omega_{fs}$ in (\ref{eqn:Darcy}) via upscaling also omits these pore-scale kinematics, the effects of which must be incorporated via coarse-grained models of, e.g., solute mixing, dispersion and reaction. Hence homogeneous Darcy flow (i.e. $\mathbf{K}(\mathbf{x})= \text{const.}$) is non-chaotic at the Darcy scale while the pore-scale flow is chaotic. In this study we focus solely on heterogeneous conductivity fields as these represent the majority of porous media~\citep{Bear:1972aa,Cushman:2013aa}.

\subsection{Kinematics of Heterogeneous Darcy Flow}

For most porous materials the hydraulic conductivity tensor $\mathbf{K}(\mathbf{x})$ is anisotropic due to anisotropy of the underlying pore-scale structure, which occurs in, e.g., stratified formations in geological media and structured engineered media~\citep{Bear:1972aa}. Under conventional upscaling approaches the conductivity tensor $\mathbf{K}(\mathbf{x})$ (a second rank tensor) is symmetric and positive definite~\citep{Bear:1972aa}, with six independent components in 3D space
 which vary spatially in heterogeneous formations but are all assumed to be smooth, continuous and finite. Hence $\mathbf{K}(\mathbf{x})$ may be locally diagonalized into principal directions with three independent components as $\mathbf{K}^\prime(\mathbf{x})=\mathbf{R}(\mathbf{x})\cdot\mathbf{K}(\mathbf{x})\cdot\mathbf{R}^{-1}(\mathbf{x})$, where $\mathbf{R}(\mathbf{x})$ is a rotation tensor field and
\begin{equation}
\mathbf{K}^\prime(\mathbf{x})\equiv\left(
    \begin{array}{ccc}
         K_{11}^\prime(\mathbf{x}) & 0 & 0\\
         0 & K_{22}^\prime(\mathbf{x}) & 0\\
         0 & 0 & K_{33}^\prime(\mathbf{x})
    \end{array}
    \right).\label{eqn:aniso}
\end{equation}
Even if $\mathbf{R}(\mathbf{x})$ is spatially invariant, this conductivity field is fundamentally anisotropic if one of the diagonal components $K_{ii}^\prime(\mathbf{x})$ differs from the others.

Conversely, many studies~\citep{Beaudoin:2013aa,Boon:2016aa,Chaudhuri:2005aa,Dartois:2018aa,Delgado:2007aa,Gelhar:1983aa,Jankovic:2009aa,Neuman:1990aa}, 
consider the hydraulic conductivity tensor to be isotropic as $\mathbf{K}(\mathbf{x})=k(\mathbf{x})\mathbf{I}$, where $k(\mathbf{x})=K_{ii}^\prime(\mathbf{x})$ for $i=1:3$, and (\ref{eqn:Darcy}) simplifies to
\begin{equation}
   \mathbf{v}(\mathbf{x})=-k(\mathbf{x})\nabla\phi,\quad \nabla\cdot\mathbf{v}=0,\,\,\,\,\mathbf{x}\in\Omega.\label{eqn:Darcy_scalar}
\end{equation}
The kinematics of isotropic Darcy flow markedly differ from that of anisotropic Darcy flow. If $k(\mathbf{x})$ is smooth and $\mathbf{v}$ does not admit stagnation points, then isotropic Darcy flow only admits simple kinematics because the helicity density $\mathcal{H}(\mathbf{x})\equiv\mathbf{v}\cdot(\nabla\times\mathbf{v})$ is identically zero
\begin{equation}
\mathcal{H}(\mathbf{x})=k\nabla\phi\cdot(\nabla\phi\times\nabla k)=0,\label{eqn:helicity}
\end{equation}
regardless of the heterogeneity of $k(\mathbf{x})$. The total helicity $H\equiv\int_\Omega\mathcal{H}(\mathbf{x})d\mathbf{x}$ over the domain $\Omega$ is a topological invariant that characterises the topological complexity (knottedness) of vortex lines of the flow~\citep{Woltjer:1958aa,Moreau:1961aa,Moffatt:1969aa}, and the helicity-free condition $H=0$ precludes chaotic dynamics in steady 3D flows~\citep{Arnold:1965aa}. Hence steady 3D isotropic Darcy flows are non-chaotic and \emph{integrable} and so admit two invariants $\psi_1$, $\psi_2$ that act as orthogonal streamfunctions~\citep{Lester:2022aa,Yoshida:2009aa}, leading to the Euler velocity representation
\begin{equation}
\mathbf{v}(\mathbf{x})=\nabla\psi_1(\mathbf{x})\times\nabla\psi_2(\mathbf{x}),\quad \nabla\psi_1(\mathbf{x})\cdot\nabla\psi_2(\mathbf{x})=0.\label{eqn:streams}
\end{equation}
Streamlines of isotropic Darcy flow are confined to the intersections of streamsurfaces given by level sets of $\psi_1(\mathbf{x})$, $\psi_2(\mathbf{x})$. Such confinement prohibits transverse dispersion in the purely advective limit $Pe\rightarrow\infty$ (henceforth simply termed transverse dispersion), regardless of medium heterogeneity~\citep{Lester:2023aa}. 

\begin{figure}
\begin{centering}
\begin{tabular}{c c c}
\includegraphics[width=0.28\columnwidth]{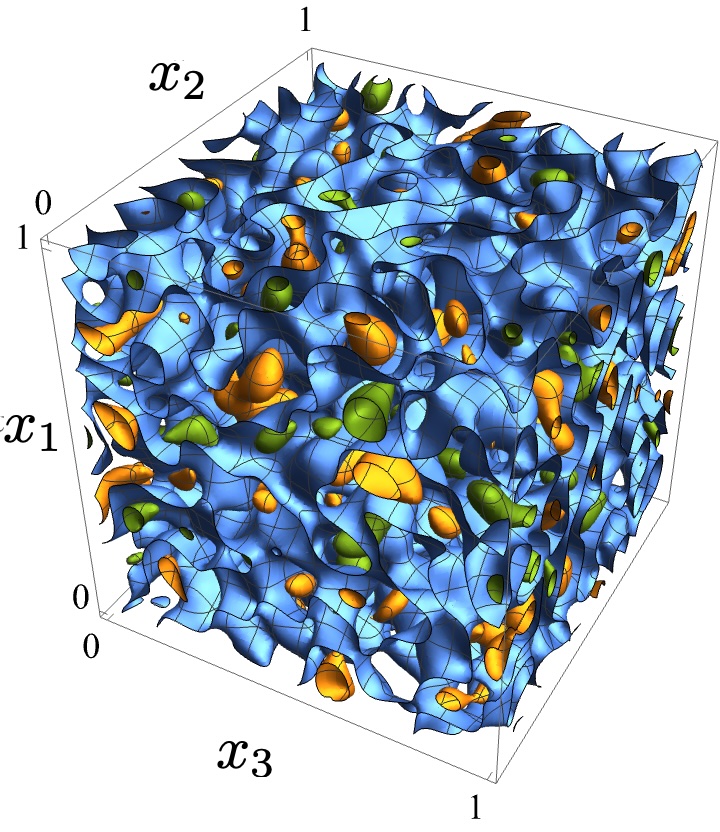}&
\includegraphics[width=0.28\columnwidth]{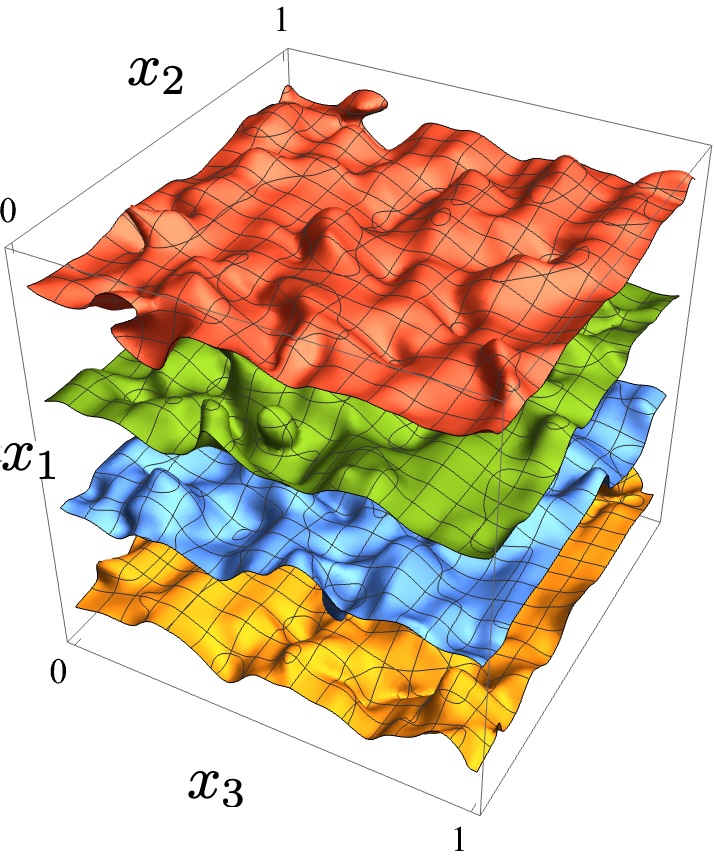}&
\includegraphics[width=0.38\columnwidth]{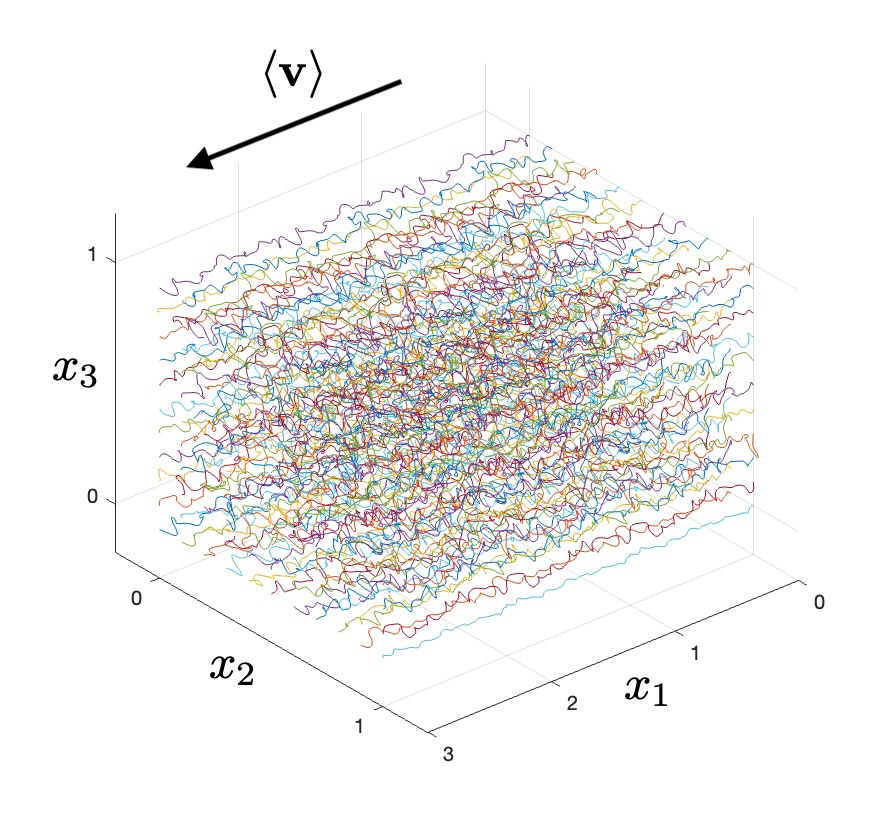}\\
(a) & (b) & (c)\\
\multicolumn{3}{c}{\includegraphics[width=0.75\columnwidth]{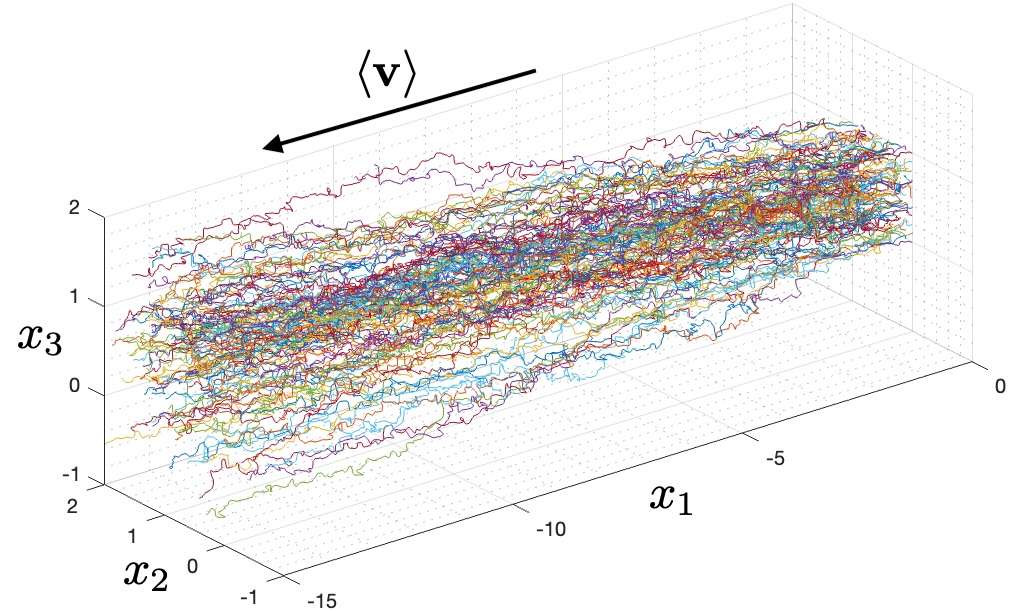}}\\
\multicolumn{3}{c}{(d)} 
\end{tabular}
\end{centering}
\caption{(a) Isosurfaces of the typical normalised heterogeneous log-conductivity field $f(\mathbf{x})=\ln K(\mathbf{x})/\sigma^2_{\ln K}$ used to model isotropic $k(\mathbf{x})\mathbf{I}$ and anisotropic $\mathbf{K}(\mathbf{x})$ conductivity tensors and (b) associated potential field $\phi(\mathbf{x})$ for anisotropic Darcy flow driven by a uniform mean potential gradient. Associated streamlines for heterogeneous Darcy flow with (c) isotropic conductivity field ($\delta=0$ in (\ref{eqn:perturb})) and (d) anisotropic conductivity field ($\delta=1$ in (\ref{eqn:perturb})) with log-conductivity variance $\sigma^2_{\ln K}=4$ and parameters $N=4$, $N_i=2$ in (\ref{eqn:logKfield}).}
\label{fig:fields}
\end{figure}

These kinematic constraints are illustrated in Figure~\ref{fig:fields}c, which shows typical streamlines for isotropic Darcy flow with strongly heterogeneous ($\sigma^2_{\ln K}=4$) hydraulic conductivity. Although the streamlines are highly tortuous due to heterogeneity of the medium, they do not exhibit asymptotic transverse dispersion. In \S\ref{sec:anisotropy} the transverse dispersivity of this flow is numerically computed to be effectively zero. 

These kinematic constraints persist even if the scalar field $k(\mathbf{x})$ is \emph{statistically} anisotropic (i.e. has different correlation structures in different principal directions), as the corresponding Darcy flow is still locally isotropic and $\mathcal{H}=0$. We remark on the application of these results to the so-called ``helicity paradox''~\citep{Cirpka:2015aa} that occurs when a statistically anisotropic but locally isotropic conductivity field is upscaled from the Darcy scale to the block field scale, resulting in an anisotropic block-scale conductivity field~\citep{Bear:1972aa}. This spuriously adds degrees of freedom to the Lagrangian kinematics and permits block-scale chaotic advection where none should exist based on the fully resolved Darcy scale flow. While beyond the scope of this paper, our results highlight the need for upscaling methods that obey the appropriate kinematic constraints.

\subsection{Implications for Hydraulic Conductivity Modelling}

These results question the validity of isotropic Darcy flow (\ref{eqn:Darcy_scalar}) as a model for heterogeneous porous systems; indeed experimental observations clearly show that transverse dispersion is non-zero toward the purely advective limit $Pe\rightarrow\infty$~\citep{Delgado:2007aa,Hackert:1996aa}. Although several numerical studies~\citep{Beaudoin:2013aa,Dartois:2018aa,Jankovic:2009aa} have found that isotropic Darcy flow can generate purely advective transverse dispersion, these observations arise in systems with either non-smooth conductivity fields or impermeable inclusions, or from numerical schemes that violate the kinematic constraints associated with (\ref{eqn:streams}). Numerical schemes that implicitly enforce these constraints yield zero transverse dispersivity~\citep{Lester:2023aa}.

Conversely, anisotropic conductivity fields $\mathbf{K}(\mathbf{x})$ generate non-zero helicity density
\begin{equation}
    \mathcal{H}(\mathbf{x})=(\mathbf{K}\cdot\nabla\phi)\cdot(\nabla\times\mathbf{K}\cdot\nabla\phi)\neq 0,\label{eqn:helicity_tensor}
\end{equation}
and non-zero total helicity $H\neq 0$. Hence the streamlines of anisotropic Darcy flow are no longer constrained to the streamsurfaces of $\psi_1(\mathbf{x})$, $\psi_2(\mathbf{x})$, but rather freely wander through the medium~\citep{Lester:2023aa}, thereby giving rise to persistent transverse dispersion. This behaviour is shown in Figure~\ref{fig:fields}d, where the streamlines associated with strongly heterogeneous ($\sigma^2_{\ln K}=4$) anisotropic Darcy flow transversely spread throughout the flow domain as they are advected longitudinally. In \S\ref{sec:anisotropy} the transverse dispersivity $D_T$ of this flow is quantitatively shown to be non-zero.

Note that non-zero helicity density $\mathcal{H}(\mathbf{x})\neq 0$ is not sufficient to ensure non-zero transverse dispersivity. The classical Clebsch 3D vector field parameterization into the smooth continuous scalar fields $\alpha_1(\mathbf{x})$, $\beta_1(\mathbf{x})$, $\varphi(\mathbf{x})$ 
\begin{equation}
\mathbf{v}(\mathbf{x})=\alpha_1(\mathbf{x})\nabla\beta_1(\mathbf{x})+\nabla\varphi(\mathbf{x})\quad\mathbf{x}\in\Omega,\label{eqn:Clebsch}
\end{equation}
has been found to be incomplete~\citep{Yoshida:2009aa}, meaning that some 3D vector fields cannot be represented by (\ref{eqn:Clebsch}). \citet{Yoshida:2017aa} classify flows conforming to (\ref{eqn:Clebsch}) as \emph{epi-2D} flows (topologically 2D) with non-zero helicity density $\mathcal{H}(\mathbf{x})\neq 0$ but are helicity free, $H=0$. Conversely, only 3D velocity fields given by the complete Clebsch parameterization~\citep{Yoshida:2009aa}
\begin{equation}
\mathbf{v}(\mathbf{x})=\sum_{i=1}^2 \alpha_i(\mathbf{x})\nabla\beta_i(\mathbf{x})+\nabla\varphi(\mathbf{x}),\label{eqn:Clebsch_comp}
\end{equation}
 have non-zero total helicity $H$ and so admit chaotic advection. Conversely, realistic heterogenenous conductivity models yield 3D velocity fields of the form (\ref{eqn:Clebsch_comp}) that admit chaotic advection and transverse dispersion.

To summarise, realistic hydraulic conductivity models of heterogeneous porous media must be anisotropic to be consistent with experimental observations~\citep{Delgado:2007aa,Hackert:1996aa} of non-zero transverse dispersion in the purely advective limit $Pe\rightarrow\infty$. As non-zero transverse dispersion requires $H\neq 0$, this raises the potential for chaotic advection, which is inevitable in nonlinear continuous systems with arbitrary coefficients and three degrees of freedom (dof=3)~\citep{Speetjens:2021aa}. In the following section we firmly establish that transverse dispersion and chaotic advection are intimately linked in heterogeneous Darcy flow.

\begin{figure}
\begin{centering}
\begin{tabular}{c c c}
\includegraphics[width=0.44\columnwidth]{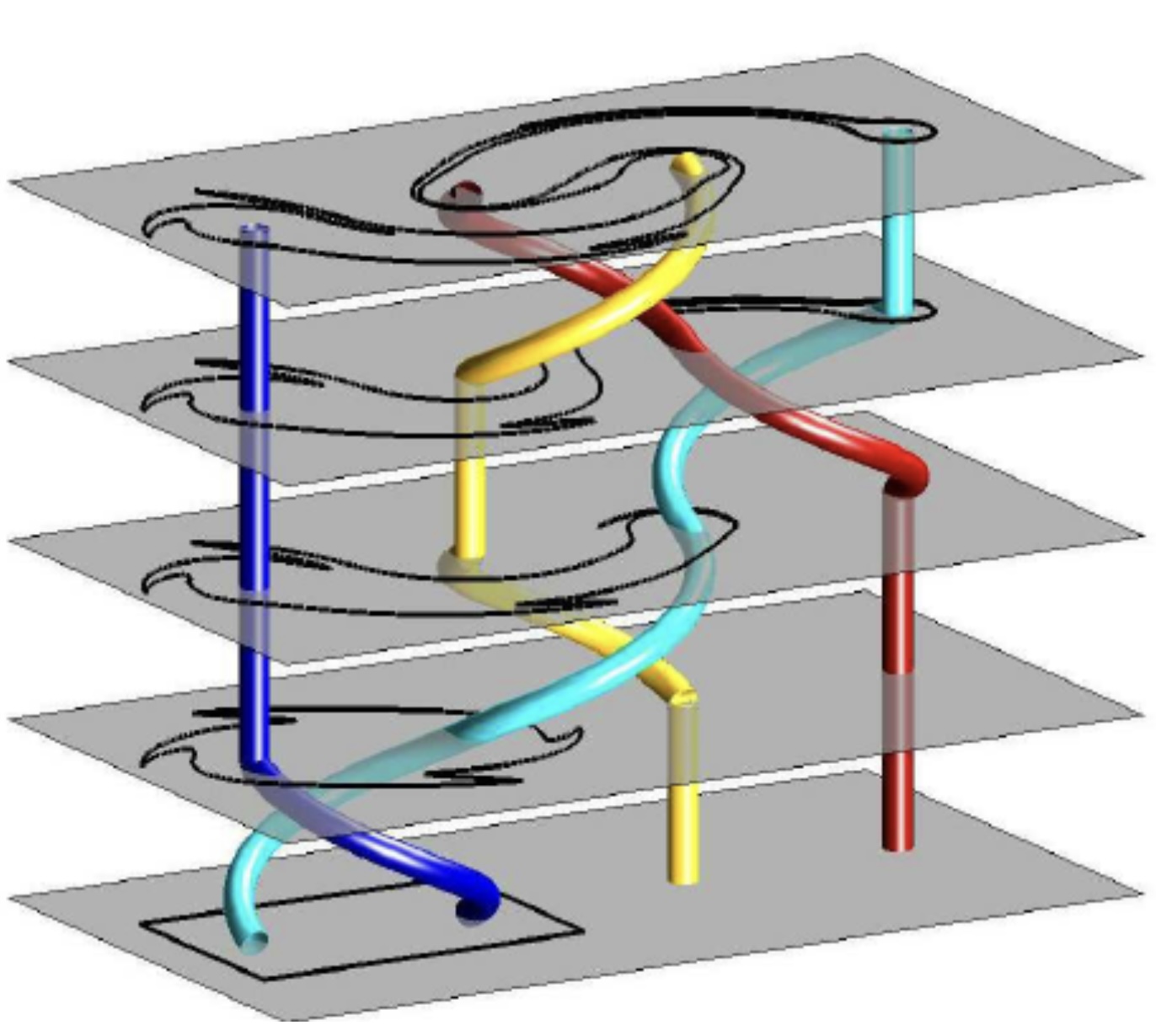}&
\hspace{0.8cm}
\includegraphics[width=0.44\columnwidth]{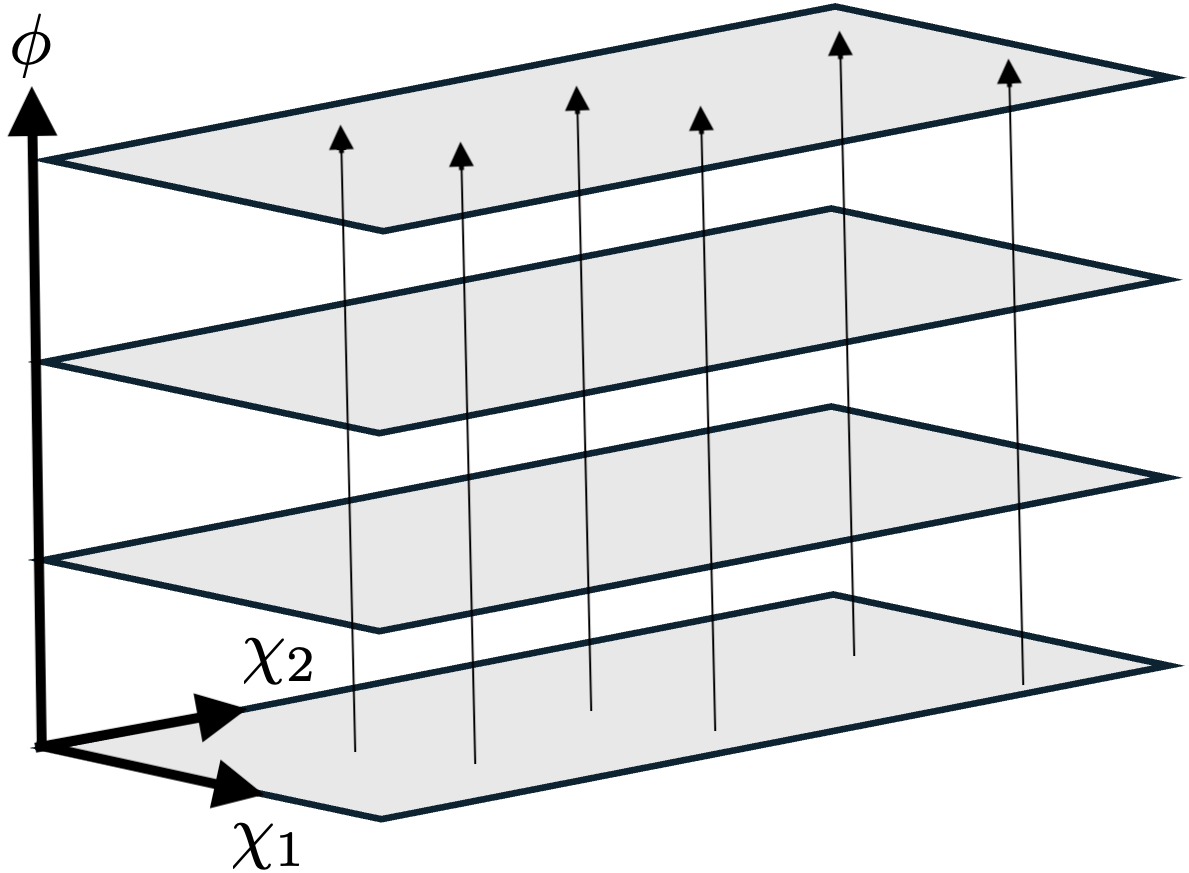}\\
(a) & (b)
\end{tabular}
\end{centering}
\caption{(a) Schematic of streamline (or pathline) braiding in a unidirectional steady 3D flow (or unsteady 2D flow). The set of four thick coloured streamlines (or pathlines) braid with each other as they evolve with the mean flow direction $x_3$ (or time $t$). This topological braiding motion stirs the fluid continuum (grey planes) and the length of the black rectangular material boundary grows exponentially with the number of braiding motions. Adapted from \citet{Thiffeault:2006aa}. This schematic also depicts streamline braiding in steady 3D anisotropic Darcy flow with intrinsic coordinates $\boldsymbol\xi=(\chi_1,\chi_2,\phi)$, where the grey planes depict isopotential surfaces.  (b) Absence of braiding in isotropic Darcy flow in intrinsic coordinates $\boldsymbol\xi=(\chi_1,\chi_2,\phi)$. As the velocity field is everywhere orthogonal to level sets of $\phi$ denoted by grey planes, streamlines in this coordinate system do not move laterally or undergo braiding.}
\label{fig:braid_schematic}
\end{figure}

\section{Streamline Braiding and Chaos}\label{sec:braiding}

\subsection{Topology of 2D Unsteady and 3D Steady Flows}\label{sec:topology}

The link between dispersion and chaotic mixing can be explored via consideration of streamline braiding. As 1D streamlines are invariants of steady 3D flow, \emph{braiding} of streamlines stirs the fluid continuum in a complex manner, as shown in Figure~\ref{fig:braid_schematic}a. This concept has been used to quantify stirring in another 3 dof system -- unsteady 2D flow -- where non-trivial braiding of 1D pathlines can stretch and fold the fluid continuum as shown in Figure~\ref{fig:braid_schematic}a. For unsteady 2D flow, braid group theory~\citep{Boyland:2000aa,Handel:1985aa,Moussafir:2006aa} has been developed to quantify the topological entropy $h_\text{braid}$ of the path-line braiding motions, closely related to the Lyapunov exponent $\lambda_\infty$~\citep{Thiffeault:2005aa}. Below we show that 2D unsteady flow is topologically equivalent to steady 3D Darcy flow, hence this mathematical framework applies to these flows.

The topology of steady 3D flow $\mathbf{v}(\mathbf{x})=[v_1(\mathbf{x}),v_2(\mathbf{x}),v_3(\mathbf{x})]$ with one unidirectional velocity component (e.g., $v_3(\mathbf{x})>0$) can be equated~\citep{Bajer:1994aa}  to that of unsteady 2D flow via the rescaling $\mathbf{v}^\prime(\mathbf{x})=\mathbf{v}(\mathbf{x})/v_3(\mathbf{x})$ such that the $x_3$ coordinate acts as an analogue for time. For strongly heterogeneous 3D Darcy flows this transformation breaks down when $v_3(\mathbf{x})$ is zero or negative due to flow reversal in the vicinity of low permeability structures. However, these flows are still fundamentally unidirectional as the potential field $\phi(\mathbf{x})$ is strictly monotonic decreasing along streamlines
$\nabla\phi(\mathbf{x})\cdot\mathbf{v}(\mathbf{x})=-\nabla\phi(\mathbf{x})\cdot\mathbf{K}(\mathbf{x})\cdot\nabla\phi(\mathbf{x})<0$
due to the positive definite nature of $\mathbf{K}(\mathbf{x})$~\citep{Bear:1972aa}. This
leads to the intrinsic coordinate basis $\boldsymbol\xi=(\chi_1,\chi_2,\phi)$, where $\chi_1$, $\chi_2$ are orthogonal to $\phi$, and the representation $
\mathbf{v}_\xi(\boldsymbol\xi)=[v_{\chi_1}(\boldsymbol\xi),v_{\chi_2}(\boldsymbol\xi),v_{\phi}(\boldsymbol\xi)]$ with $v_{\phi}(\boldsymbol\xi)>0$. Thus, the velocity field $\mathbf{v}_\xi(\boldsymbol{\xi})$ may be rescaled as
\begin{equation}
\mathbf{v}_\xi^\prime(\boldsymbol\xi)=\left[\frac{v_{\chi_1}(\boldsymbol\xi)}{v_{\phi}(\boldsymbol\xi)},\frac{v_{\chi_2}(\boldsymbol\xi)}{v_{\phi}(\boldsymbol\xi)},1\right],\quad v_{\phi}(\boldsymbol\xi)>0\,\forall\,\boldsymbol\xi\in\Omega.\label{eqn:vxi_prime}
\end{equation}
This velocity field is topologically equivalent to that of unsteady 2D flow, and the isopotential surfaces of steady 3D Darcy flow (Figure~\ref{fig:fields}b) are analogous to isochrones of unsteady 2D flow. Hence streamlines in steady 3D Darcy flow are topologically equivalent to pathlines of unsteady 2D flow and the mathematical framework developed for pathline braiding can be directly applied to steady 3D Darcy flow. As per Figure~\ref{fig:braid_schematic}b, the representation (\ref{eqn:vxi_prime}) also shows directly that isotropic Darcy flow cannot generate streamline braiding, where $(\chi_1,\chi_2)=(\psi_1,\psi_2)$ and (\ref{eqn:vxi_prime}) simplifies to $\mathbf{v}_\xi^\prime(\boldsymbol\xi)=(0,0,1)$ because the velocity field is everywhere orthogonal to the level sets of $\phi$, i.e. $\mathbf{v}\times\nabla\phi=\mathbf{0}$. These streamlines do not move relative to each other in the isopotential surfaces $\phi= \text{const.}$, let alone undergo non-trivial braiding motions.  In Cartesian coordinates (Figure~\ref{fig:fields}c), the helicity-free condition constrains non-braiding yet tortuous streamlines to streamsurfaces $\psi_1(\mathbf{x})$=const., $\psi_2(\mathbf{x})$=const.

Conversely, anisotropic Darcy flow admits non-zero transverse velocity components $v_{\chi_1}(\boldsymbol\xi)$, $v_{\chi_2}(\boldsymbol\xi)$ and so streamlines can undergo the braiding motions as shown in Figure~\ref{fig:braid_schematic}a and Figure~\ref{fig:fields}d. Although transverse motion of streamlines in the $(\chi_1,\chi_2)$ directions does not necessarily generate non-trivial braiding, below we show that non-trivial braiding is inherent to steady random 3D flows.

\begin{figure}
\begin{centering}
\begin{tabular}{c c c}
\multicolumn{2}{c}{\includegraphics[width=0.7\columnwidth]{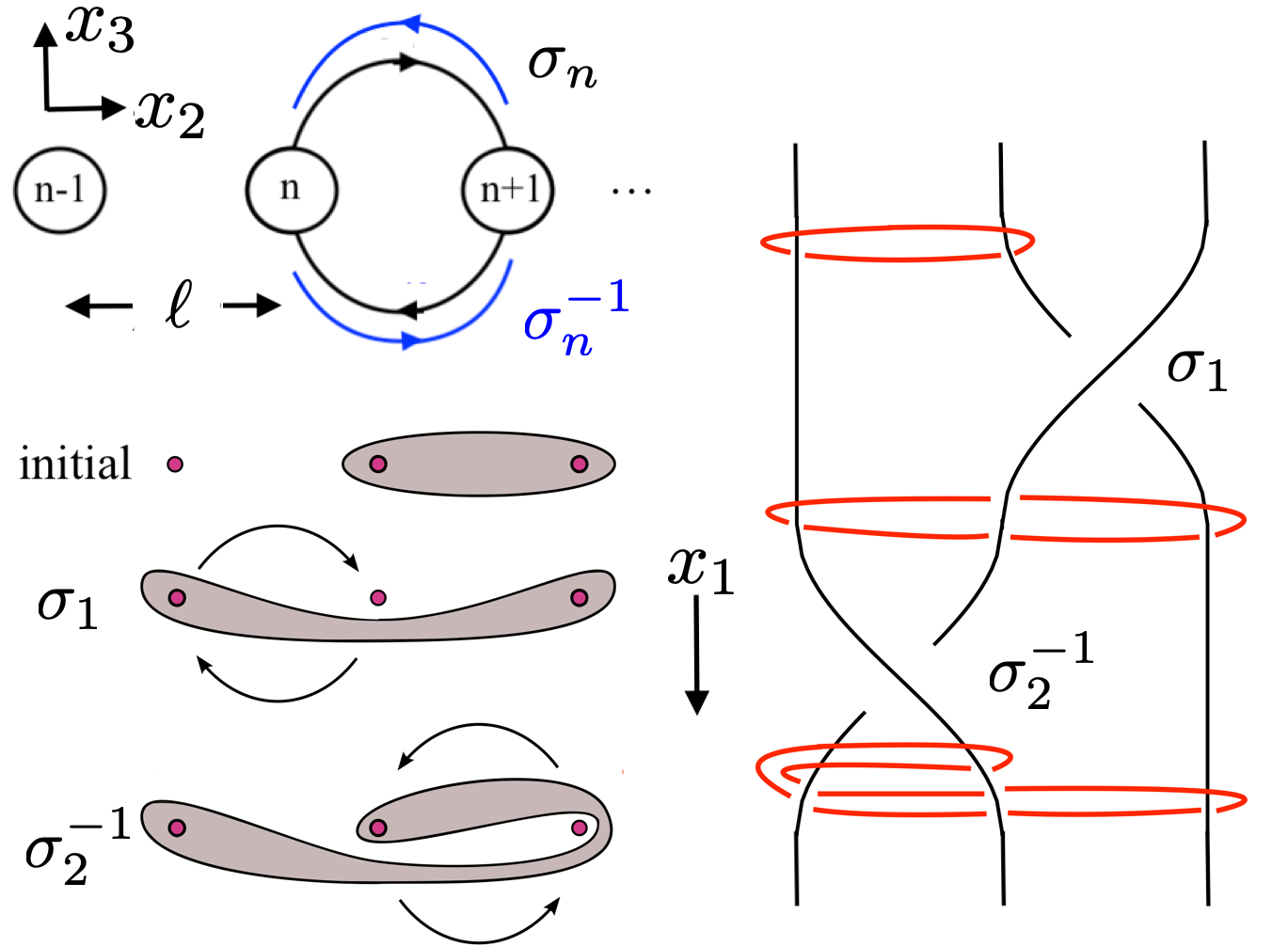}}\\
\multicolumn{2}{c}{(a)}\\
\includegraphics[width=0.43\columnwidth]{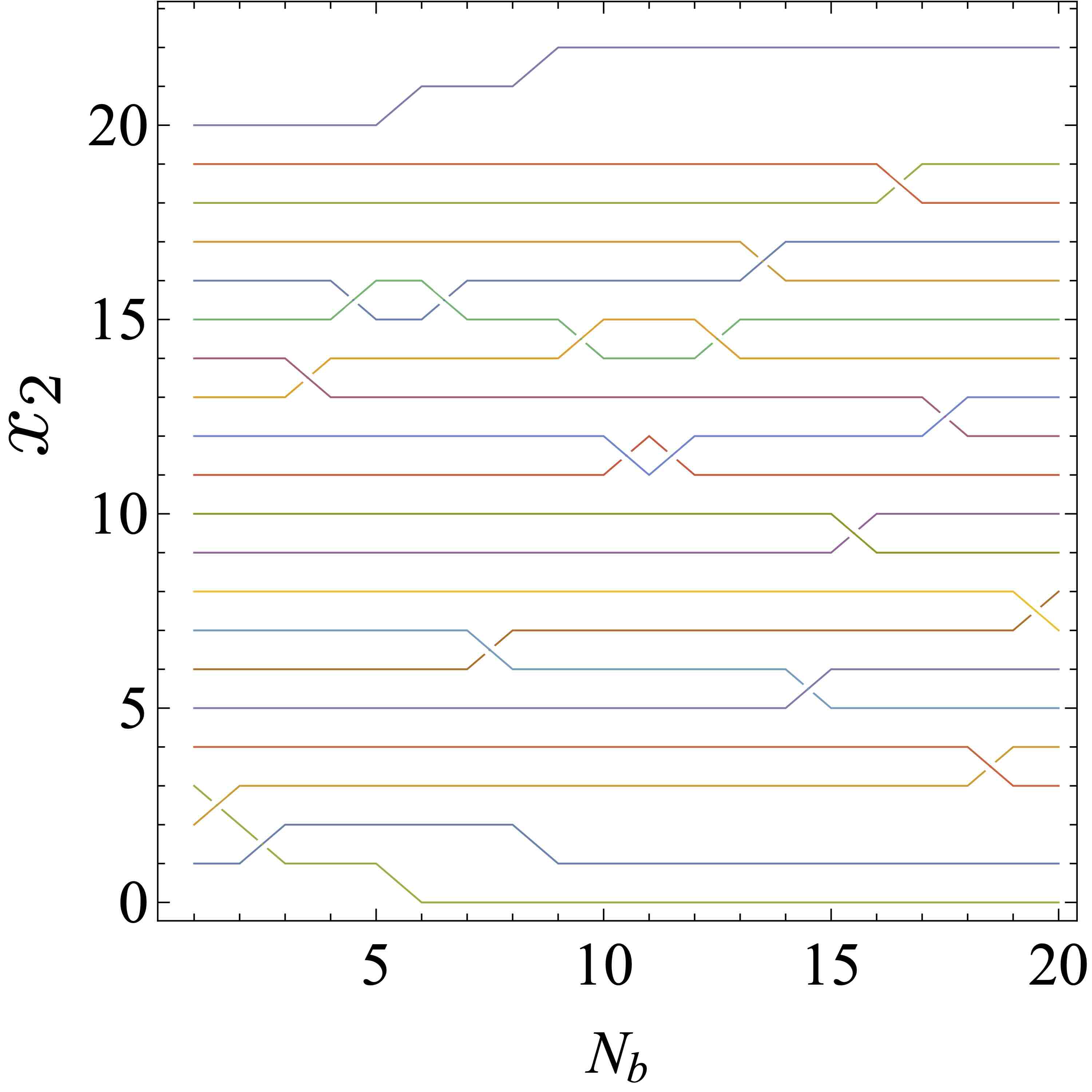}&
\includegraphics[width=0.45\columnwidth]{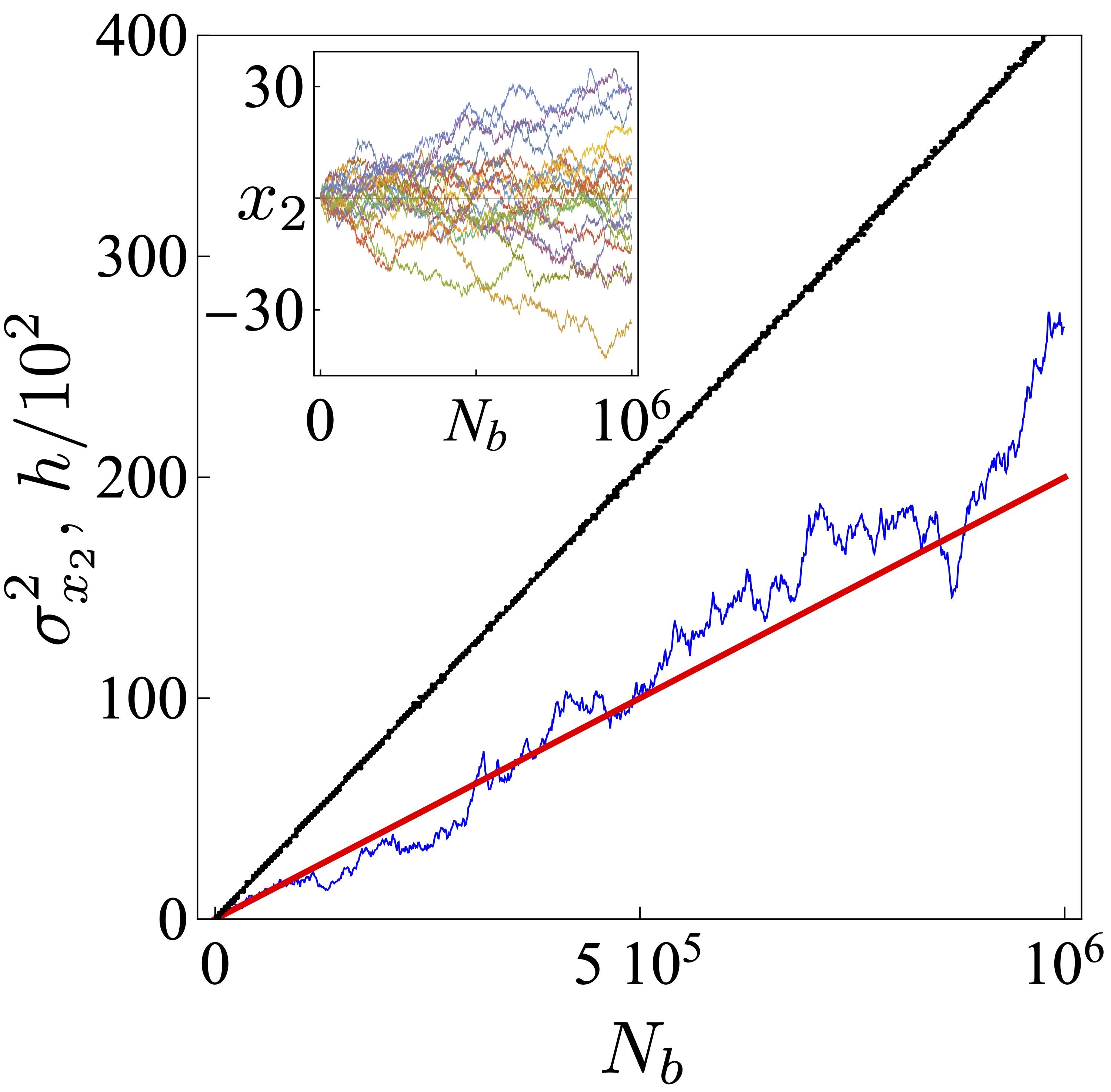}\\
 (b) & (c)
\end{tabular}
\end{centering}
\caption{(a) Top left: schematic of streamlines $n-1$, $n$, $n+1$ in the transverse $x_2-x_3$ plane with clockwise (black) $\sigma_n$, anti-clockwise (blue) $\sigma_n^{-1}$ braid generators in the 1D streamline model. Bottom left: stretching of material elements due to braiding motions (adapted from \citet{braidbook}) that evolve in the longitudinal $\phi$ direction. Right: braid diagram (black) depicting stretching of material elements (red) as braid evolves in longitudinal $\phi$ direction. 
(b) Braid diagram depicting evolution of $x_2$ coordinate of $N_p=20$ streamlines over $N_b=20$ random braid actions in the longitudinal $\phi$ direction, leading to non-trivial braiding and dispersion of streamlines. (c) Growth of topological entropy $h$ (black line) and transverse variance $\sigma_{x_2}^2$ (blue line) with braid number $N_b$, which agrees well with theory~\citep{Lester:2024aa} (red line). Inset: Brownian motion of streamlines due to streamline braiding with braid number $N_b$. Adapted from \citet{Lester:2024aa}.
}
\label{fig:braid}
\end{figure}

\subsection{Topological Complexity of 3D Streamline Braiding}\label{subsec:topological}

Braid group theory is used to quantify stirring in unsteady 2D flows~\citep{Boyland:2000aa} via a well-developed mathematical framework~\citep{Artin:1947aa,Moussafir:2006aa, braidbook} that efficiently encodes fluid stirring via a symbolic braiding representation of the topology of pathlines in spacetime. This encoding can be used to compute the topological complexity of their braiding motions, and the Neilson-Thurston classification theorem~\citep{Handel:1985aa} characterises braids as periodic, reducible or pseudo-Anosov types, the latter of which is interpreted as an indicator of chaotic dynamics~\citep{Boyland:2000aa}. Due to topological equivalence, this framework can also be applied to streamline braiding in steady 3D Darcy flow.

For steady unidirectional 3D flow, streamline braiding is encoded via a sequence of \emph{braid generators} $\sigma_n^{\pm 1}$~\citep{Artin:1947aa}. As shown in Figure~\ref{fig:braid}a, braiding of streamlines (circles) in the $(x_2,x_3)$ plane of $\mathbb{R}^3$ space can be characterised in terms of the ``crossing'' of these streamlines with respect to arbitrary reference direction $x_3$ as they propagate unidirectionally in the $x_1$ direction (into the page in Figure~\ref{fig:braid}a). Note streamlines are labelled ($\dots,n-1, n,n+1,\dots$) with respect to their relative position in the $x_2$ coordinate, hence these labels are updated after each crossing event. The braid generators $\sigma_n$ and $\sigma_n^{-1}$ respectively characterise clockwise and counter-clockwise crossing of streamlines $n$ and $n+1$. An ordered sequence of braid generators form the \emph{braid word} $\mathbf{b}=\sigma_{n_1}^{\pm1}\sigma_{n_2}^{\pm1}\dots$, which is determined by the sequence of streamline crossings in the $x_1$ direction. 
As shown in Figure~\ref{fig:braid}a (bottom left and right), these braiding motions act to stretch and fold fluid elements (brown), leading to exponential stretching if they are of pseudo-Anosov type.

The topological entropy $\hat{h}$ of a dynamical system measures the rate of loss of information of the system about its initial conditions. For fluid flows $\hat{h}$ is closely related to the largest infinite-time Lyapunov exponent $\hat{\lambda}_\infty$ and forms an upper bound for $\hat{\lambda}_\infty$~\citep{Thiffeault:2010aa}. In practice, the topological entropy of a flow can be interpreted as the asymptotic exponential growth rate of the length $l(t)$ of a material line~\citep{Newhouse:1993aa} as
\begin{equation}
\hat{h}=\lim_{t\rightarrow\infty}\frac{1}{t}\ln \frac{l(t)}{l(0)}.\label{eqn:top_entropy}
\end{equation}
For ergodic flows such as purely hyperbolic steady 3D flow (which has a single positive Lyapunov exponent), the upper bound given by the topological entropy is exact~\citep{Yomdin:1987aa,Newhouse:1988aa, Newhouse:1993aa, Matsuoka:2015aa,Catalan:2019aa}, i.e.
\begin{equation}
\hat{h}=\hat{\lambda}_\infty.\label{eqn:entropy_mean}
\end{equation}
Conversely, many studies dating over half a century~\citep{Villermaux:2019aa, Duplat:2010ab, Kalda:2000aa, Hinch:1999aa, Girimaji:1990aa,Cocke:1969aa} propose that as fluid elements undergo stretching, the relative length $\rho\equiv\delta l(t)/\delta l(0)$ of infinitesimal line elements is distributed log-normally with log-mean $\hat{\lambda}_\infty t$ and log-variance $\hat{\sigma}^2_\lambda t$. In \S\ref{sec:heterogeneity} and Appendix~\ref{app:1Dstretch} we confirm that an \emph{ab initio} derivation of fluid stretching in steady 3D flows yields the same result if the velocity distribution does not yield anomalous transport. For this scenario a reasonable conclusion~\citep{Villermaux:2019aa,Duplat:2010ab,Hinch:1999aa} is that the growth rate of $l(t)=\int_0^{l(0)}\delta l(t)/\delta l(0) dl$ is given by the ensemble average of $\rho$, which, via (\ref{eqn:top_entropy}), yields
\begin{equation}
\hat{h}=\hat{\lambda}_\infty+\frac{\sigma^2_\lambda}{2}.\label{eqn:entropy_var}
\end{equation}
as the variance of $\ln\rho$ contributes to the growth of $\ln l(t)$. However, the discrepancy of (\ref{eqn:entropy_var}) with (\ref{eqn:entropy_mean}) has recently been explained by \citet{Lester:2024ab}, who show that in the asymptotic limit, temporal sampling of the finite-time Lyapunov exponent $\hat{\lambda}(\mathbf{X},t)$ (which converges to $\hat{\lambda}_\infty$ as $\sigma_\lambda/\sqrt{t}$) dominates over the ensemble average represented by (\ref{eqn:entropy_var}), yielding (\ref{eqn:entropy_mean}). In \S\ref{sec:mechanisms} we show that heterogeneous Darcy flow with random stationary conductivity $\mathbf{K}(\mathbf{x})$ is ergodic and purely hyperbolic (i.e. it does not admit non-hyperbolic features such as elliptic or parabolic Lagrangian Coherent Structures). 

The dimensionless topological entropy $h\equiv \hat{h}\ell/\langle v_1\rangle$ (with velocity correlation length  $\ell$) may be approximated by considering the topological entropy $h_\text{braid}$ of the braid word $\mathbf{b}$ of a set of streamlines of the flow. The braid entropy $h_\text{braid}$ characterises the asymptotic growth (with number of braiding motions $k=t\langle v_1\rangle/\ell$) of the number of \emph{distinguishable orbits} of the braid~\citep{braidbook}. For steady 3D flow this may be interpreted as the number of distinct linkages that a material line (e.g. Figure~\ref{fig:braid}a) makes if it is fully contracted until it forms a \emph{loop} $\ell_E$ that tightly contacts these streamlines. The topological entropy $h_\text{braid}$ of a braid word $\mathbf{b}$ is then
\begin{equation}
h_\text{braid}=\lim_{k\rightarrow\infty}\frac{1}{k}\log \frac{|\mathbf{b}^k \ell_E|}{|\ell_E|},
\end{equation}
where the metric $|\cdot|$ measures the number of linkages of the braid. Although non-trivial, various methods~\citep{Moussafir:2006aa,Hall:2009aa,braidbook} are available to rigorously compute $h_\text{braid}$ from a given braid word $\mathbf{b}$. As the braid entropy $h_\text{braid}$ only considers the complexity of streamline braiding, it forms a lower bound for $h$, however for many systems it converges to $h$ with increasing number $N_p$ of streamlines as
\begin{equation}
\lim_{N_p\rightarrow\infty}h_\text{braid}\rightarrow h=\lambda_\infty.
\end{equation}
Hence for heterogeneous Darcy flow both the topological entropy $h$ and dimensionless Lyapunov exponent $\lambda_\infty\equiv \hat{\lambda}_\infty\ell/\langle v_1\rangle$ may be accurately estimated from the braid entropy $h_\text{braid}$ of a sufficiently large set of streamlines.

\subsection{Linking Dispersion and Chaos in Streamline Braiding Flows}

\citet{Lester:2024aa} recently examined the link between chaotic advection and transverse dispersion in randomly braiding flows via a simple 1D streamline model. As per Figure~\ref{fig:braid}a~(top left), this model consists of a set of $N_p$ streamlines in $\mathbb{R}^3$ space that repeat periodically in the transverse $x_2$ direction and propagate due to mean flow in the longitudinal $x_1$ direction. These streamlines have the same $x_3$-coordinate ($x_3=0$) and uniform spacing $\Delta x_2=\ell$ in the $x_2$-direction, corresponding to the velocity correlation length scale $\ell$. At integer multiples of longitudinal distance $\Delta_L$, one pair ($n$, $n+1$) of neighbouring streamlines randomly exchange position in the $x_2-x_3$ plane via clockwise or counter-clockwise rotations, as labelled by the respective braid generators $\sigma_n$, $\sigma_n^{-1}$.  

The streamline braiding motions are defined by a braid word $\mathbf{b}$ comprised of $N_b$ braid generators $\mathbf{b}=\sigma_{n_1}^{\pm1}\sigma_{n_2}^{\pm1}\dots\sigma_{n_{N_b}}^{\pm1}$ which is constructed by randomly choosing each generator from the set $\{\sigma_1,\sigma_1^{-1},\dots,\sigma_{N_p},\sigma_{N_p}^{-1}\}$. Streamlines undertake an unbounded random walk along the $x_2$ coordinate as they propagate longitudinally (Figure~\ref{fig:braid}c, inset), while deforming the interstitial fluid (Figure~\ref{fig:braid}a, right). The topological braid entropy $h_\text{braid}$ of these motions is efficiently computed via the \emph{braidlab} package~\citep{braidlab}. Figure~\ref{fig:braid}b shows the braid diagram for a typical braid word $\mathbf{b}$ consisting of $N_p=20$ streamlines and $N_b=20$ braid generators. Figure~\ref{fig:braid}c shows linear growth of topological braid entropy $h_\text{braid}$ and transverse variance $\sigma_{x_2}^2$ of this system with $N_b$.

\citet{Lester:2024aa} show that the topological braid entropy $h_\text{braid}$ and transverse variance $\sigma_{x_2}^2$ of this system grow linearly with $N_b$, and in the limit of large $N_b$ and $N_p$, the scaled mean topological braid entropy $\langle h_\text{braid}\rangle$ for $10^3$ realizations of this simple 1D streamline model converges to the \emph{random braid entropy} $\langle\lambda_\sigma\rangle$ as
\begin{equation}
\lim_{N_p\rightarrow\infty}\lim_{N_p\rightarrow\infty}\langle h_\text{braid}\rangle \frac{N_p}{N_b}\rightarrow\langle\lambda_\sigma\rangle\approx 0.8529,  
\end{equation}
leading to a simple relationship between $\lambda_\infty$ and the transverse dispersivity $D_T\equiv\lim_{t\rightarrow\infty}d/dt\,\sigma^2_{x_2}/2$ as
 \begin{equation}
\lambda_\infty=\frac{\langle\lambda_\sigma\rangle}{Pe_T},\label{eqn:model1D}
\end{equation}
 where $Pe_T\equiv\ell \langle v_1\rangle/D_T$ is the transverse P\'{e}clet number.
 
Two different 2D extensions of this 1D streamline model were also considered by \citet{Lester:2024aa}; one involving a 2D streamline array analogous to the 1D array in Figure~\ref{fig:braid}a, and another involving 3D streamlines that undertake independent random walks constructed by making a jump of fixed magnitude by random orientation in the $x_2-x_3$ plane at regular intervals in the $x_1$ coordinate. For both of these models the appropriately scaled topological braid entropy $h_\text{braid}$ was also found to converge to the random braid entropy $\langle\lambda_\sigma\rangle$, leading to a simple relationship between the Lyapunov exponent $\lambda_\infty$ of the flow and the transverse P\'{e}clet number $Pe_T$ as
\begin{equation}
\lambda_\infty^d=\langle\lambda_\sigma\rangle^d\left(\frac{\ell}{\Delta_L}\right)^{d-1}\frac{d}{Pe_T},\label{eqn:model}
\end{equation}
where $d=1,2$ is the Euclidean dimension of the streamline array. The persistence of this relationship across these diverse models suggests that they all belong to the same \emph{universality class}~\citep{Odor:2004aa} associated with streamline braiding. This establishes that chaotic advection and transverse dispersion are intimately linked in heterogeneous Darcy flow because they are both driven by non-trivial streamline braiding. Equation (\ref{eqn:model}) also points to development of methods to estimate $\lambda_\infty$ from transverse dispersivity data for specific systems, however further research is required to extend this link to non-ideal systems.

\subsection{Fluid Deformation in Heterogeneous Darcy Flow}\label{subsec:deform}

In addition to streamline braiding, it is also instructive to consider \emph{ab initio} evolution of fluid deformation in heterogeneous Darcy flow from a kinematic perspective. This can be quantified by considering deformation as a random process. \citet{Le-Borgne:2008aa,Le-Borgne:2008ab} established that for steady flow in random media, the velocity magnitude $v$ decorrelates with distance $s$ along streamlines and is described by a spatial Markov process with respect to spatial correlation length $\ell$. It has also been shown~\citep{Lester:2022aa} that the velocity gradient in steady Darcy flow is also spatially Markovian, hence fluid deformation is characterised by sampling the dimensionless velocity gradient tensor $\boldsymbol\epsilon(t;\mathbf{X})\equiv\nabla\mathbf{v}(\mathbf{x}(t;\mathbf{X}))^\top \ell/\langle v_1\rangle$ equidistantly with distance $s$ along streamlines. Rotation into the \emph{Protean} coordinate frame $\mathbf{x}^\prime$~\citep{Lester:2018aa} renders the velocity gradient tensor $\boldsymbol\epsilon^\prime(t;\mathbf{X})$ upper triangular (see Appendix~\ref{app:Lyapunov} for details) and the fluid deformation gradient tensor $\mathbf{F}^\prime(t;\mathbf{X})$ which evolves as
\begin{equation}
\frac{d\mathbf{F}^\prime(t;\mathbf{X}^\prime)}{dt}=\boldsymbol\epsilon^\prime(t;\mathbf{X})\cdot\mathbf{F}^\prime(t;\mathbf{X}),\quad \mathbf{F}^\prime(0;\mathbf{X})=\mathbf{1},\label{eqn:deform}
\end{equation}
is likewise. The diagonal elements $F_{ii}^\prime$ represent principal stretches and the off-diagonal components $F_{ij}^\prime$ (non-zero for $j>i$) represent shear deformations. Hence the hyperbolic stretches that govern the Lyapunov exponents are given by the diagonal components $\epsilon_{ii}^\prime$, and are not conflated with shear deformations given by the off-diagonal components $\epsilon_{ii}^\prime$. From (\ref{eqn:deform}), the diagonal elements of $\mathbf{F}^\prime(t;\mathbf{X})$ grow exponentially as
\begin{equation}
\begin{split}
F_{ii}^\prime(t;\mathbf{X}^\prime)
=\exp\left[\int_0^t dt^\prime \epsilon_{ii}^\prime(t^\prime;\mathbf{X})\right],
\quad i=1,2,3,\label{eqn:Fii}
\end{split}
\end{equation}
From (\ref{eqn:Fii}), fluid stretching in the streamwise direction simply fluctuates with the local velocity as $F_{11}^\prime(t)=v(t)/v(0)$, hence $\langle\epsilon^\prime_{11}(t)\rangle=0$. As $\sum_{i}\epsilon^\prime_{ii}=0$ for divergence free flow flows, the Lyapunov exponent is then
\begin{equation}
\lambda_\infty=\langle\epsilon^\prime_{22}\rangle=-\langle\epsilon^\prime_{33}\rangle,\label{eqn:Lyapunov}
\end{equation}
where $\langle \cdot\rangle$ denotes spatial averaging along streamlines. Hence $F_{22}^\prime\sim\exp(\lambda_\infty t)$ and $F_{33}^\prime\sim\exp(-\lambda_\infty t)$, and so chaotic advection in steady 3D flow is characterised by the single Lyapunov exponent $\lambda_\infty$.


In Appendix~\ref{app:1Dstretch} we show that (\ref{eqn:Fii}) leads to an \emph{ab initio} continuous time random walk (CTRW) for the relative stretch $\rho\equiv \delta l(\mathbf{X},t)/\delta l(\mathbf{X},0)$ of infinitesimal fluid line elements along streamlines as
\begin{align}
\begin{split}
\ln \rho_{n+1} &= \ln\rho_n + \frac{\ell\,\epsilon_n}{v_n},\\
 t_{n+1} &= t_n + \tau_n = t_n + \frac{\ell}{v_n},\\
 s_{n+1} &=s_n+\ell,\label{eqn:ctrw}
\end{split}
\end{align}
where $v_n$ and $\tau_n$ respectively are the streamline velocity magnitude and advection time between $s_n$ and $s_{n+1}$ and $\epsilon_n$ is the relevant Protean velocity gradient component $\epsilon^\prime_{22}$. The distribution $\psi(\tau)$ of waiting times $\tau$ is related to the Lagrangian velocity distribution $p_s(v)$ as
\begin{equation}
\psi(\tau)=\frac{\ell p_s(\ell/\tau)}{\tau^2},
\end{equation}
and $p_s(v)=v p_e(v)/\langle v \rangle_e$~\citep{Dentz:2016aa}, where $p_e(v)$ is the Eulerian velocity distribution and $\langle \cdot \rangle_e$ represents an Eulerian average. For heterogeneous Darcy flow the Lagrangian velocity distribution may be heavy-tailed in the small velocity limit~\citep{Berkowitz:2006aa}, i.e., $p_s(v)\propto v^{\beta-1}$, $\beta>1$ for $v\ll\langle v\rangle$ depending on the distribution of hydraulic conductivity \citep{Hakoun:2019aa}. Thus, the waiting time distribution scales as 
\begin{equation}
\psi(\tau)\approx\frac{\psi_0}{|\Gamma(-\beta)|} t^{-1-\beta}\quad\text{for}\,\,\tau/\langle \tau\rangle\gg 1,
\end{equation}
and the $q$-th order moments $\langle \tau^q\rangle$ of $\psi(\tau)$ are finite for $q<\lfloor \beta\rfloor$. Hence normal transport arises if $\beta>2$, as the mean and variance of $\psi(\tau)$ are bounded, but anomalous transport arises for $1<\beta<2$ as the variance of $\psi(\tau)$ is unbounded. For $\beta>1$, the advection time $t_n$ may be related to the number of increments $n$ as
\begin{equation}
t_n=\sum_{i=1}^n\tau_i\approx n\langle \tau\rangle = n\ell\left\langle\frac{1}{v_n}\right\rangle=\frac{n\ell}{\langle v\rangle_e}=n\tau_c,
\end{equation}
where $\tau_c\equiv\ell/\langle v\rangle_e$ is the mean transition time. Similarly, the mean of $\ln \rho$ also grows linearly in time as $\langle\ln \rho\rangle=\langle \epsilon_n\rangle t=\lambda_\infty t$.
The rate of growth of the variance $\sigma^2_{\ln\rho}$ serves as an important input parameter for lamellar mixing theories~\citep{Le-Borgne:2015aa,Villermaux:2019aa} that predict evolution of the concentration PDF for advective-diffusive solute transport from fluid stretching behaviour. \citet{Dentz:2015ab} and \citet{Rebenshtok:2014aa} show that in the asymptotic limit the coupled CTRW (\ref{eqn:ctrw}) generates $q$-th order absolute central moments $\mu_q$ for $q>\beta$ which scale as
\begin{equation}
\mu_q(t)\equiv\lim_{t\rightarrow\infty}\langle|\ln \rho-\lambda_\infty t |^q\rangle\sim
\begin{cases}
t^{1+q-\beta}\quad&\text{for}\,\,1<\beta<2,\\
t^{q/2}\quad&\text{for}\,\,\beta>2,
\end{cases}
\end{equation}
hence anomalous transport ($\beta<2$) in heterogeneous 3D Darcy flow can also generate anomalous stretching dynamics. In Appendix~\ref{app:1Dstretch} we show for the case $\beta>2$, $p_{\ln\rho}(\ln\rho)$ converges to a normal distribution with mean and variance
\begin{align}
\lim_{t\rightarrow\infty}\langle\ln\rho\rangle=\lambda_\infty t && \lim_{t\rightarrow \infty}\sigma^2_{\ln\rho}=\sigma^2_\lambda t,
\end{align}
where $\sigma^2_\lambda\equiv\sigma^2_\epsilon\langle\tau^2\rangle/\tau_c$ and $\sigma_\epsilon^2$ is the variance of $p_{\epsilon}(\epsilon)$. For the anomalous stretching case ($1<\beta<2$), the variance $\sigma^2_{\ln\rho}$ growth ranges from normal to ballistic as~\citep{Dentz:2015ab} 
\begin{align}
\lim_{t\rightarrow \infty}\sigma^2_{\ln\rho}=\frac{\psi_0\,\beta\,\Gamma(\beta+1)}{6-5\beta+\beta^2}t^{3-\beta}.
\end{align}
Under such anomalous stretching the equality (\ref{eqn:entropy_mean}) between $h$ and $\lambda_\infty$ persists as temporal averaging along streamlines still dominates~\citep{Lester:2024ab}, but convergence of (\ref{eqn:top_entropy}) slows as $\beta\downarrow 1$ and physical processes such as mixing and dispersion may be impacted in this limit. In \S\ref{sec:numerics}, we find that the flows considered herein all exhibit normal transport. However, anomalous stretching dynamics form an important kinematic regime that warrants future investigation.

\section{Chaotic Advection in Heterogeneous Darcy Flow}\label{sec:numerics}

\subsection{Onset of Chaotic Advection with Anisotropy}\label{sec:anisotropy}

The results in \S\ref{sec:braiding} uncover a deep link between chaotic advection and transverse dispersion in heterogeneous Darcy flow, and show how anomalous transport generates anomalous stretching dynamics. To examine the onset of chaotic advection with medium anisotropy, we first consider Darcy flow in the simplest possible conductivity field that admits non-zero helicity:
\begin{equation}
\mathbf{K}(\mathbf{x})=k_0(\mathbf{x})\mathbf{I}+\delta[k_\delta(\mathbf{x})-k_0(\mathbf{x})]\hat{\mathbf{e}}_1\otimes\hat{\mathbf{e}}_1,\label{eqn:perturb}
\end{equation}
where $k_0(\mathbf{x})\neq k_\delta(\mathbf{x})$ and $\delta\in[0,1]$ quantifies anisotropy of the conductivity tensor. Although this flow has been previously considered~\citep{Lester:2024aa} in the context of quantifying the link between stirring and dispersion, here we focus on the onset of chaotic dynamics with $\delta$. Note that while similar conductivity tensor fields such as $\mathbf{K}(\mathbf{x})=k_0(\mathbf{x})(\mathbf{I}+\delta\hat{\mathbf{e}}_1\otimes\hat{\mathbf{e}}_1)$ have non-zero helicity density $\mathcal{H}(\mathbf{x})\neq 0$, the resulting flows are shown (Appendix~\ref{app:zeroH} to be epi-2D~\citep{Yoshida:2017aa} and hence non-chaotic~\citep{Holm:1991aa,Arnold:1965aa}.

Darcy flow over the conductivity field (\ref{eqn:perturb}) is driven by a unit potential gradient $\nabla\bar{\phi}=\{-1,0,0\}$ in a triply-periodic unit cube (3-torus $\mathbb{T}^3$) $\Omega:\mathbf{x}\in[0,1]\times[0,1]\times[0,1]$, and the scalar log-conductivity fields $f(\mathbf{x})\equiv\ln  k(\mathbf{x})$ for the independent fields $k_0(\mathbf{x})$, $k_\delta(\mathbf{x})$ are given by
\begin{equation}
\begin{split}
f(\mathbf{x})&=\Sigma^2_{\ln K}\sum_{n=1}^{N_i}\sum_{i,j,k}^{N} \frac{A_{n,ijk}}{\sqrt{i^2+j^2+k^2}}
\cos[2\pi i (x_1+\chi^1_{n,ijk})]\\
&\cos[2\pi j (x_2+\chi^2_{n,ijk})]\cos(2\pi k (x_3+\chi^3_{n,ijk})),\label{eqn:logKfield}
\end{split}
\end{equation}
with $N=4$ and so the velocity correlation length $\ell=1/(2N)$. $N_i=2$ is the number of realisations in each mode, and $A_{n,ijk}$ and $\chi^m_{n,ijk}$ with $m=1,2,3$ are uniformly distributed random variables in $[0,1]$. The coefficient $\Sigma_{\ln K}$ is chosen such that the log-variance of the conductivity field is $||f^2||_\Omega=\sigma^2_{\ln K}=4$. A typical scalar field $f(\mathbf{x})$ with these parameters is shown in Figure~\ref{fig:fields}a. Note fields with $N=1$ do not generate chaotic dynamics due to symmetry of the velocity field.

\begin{figure}
\begin{centering}
\begin{tabular}{c c}
\includegraphics[width=0.51\columnwidth]{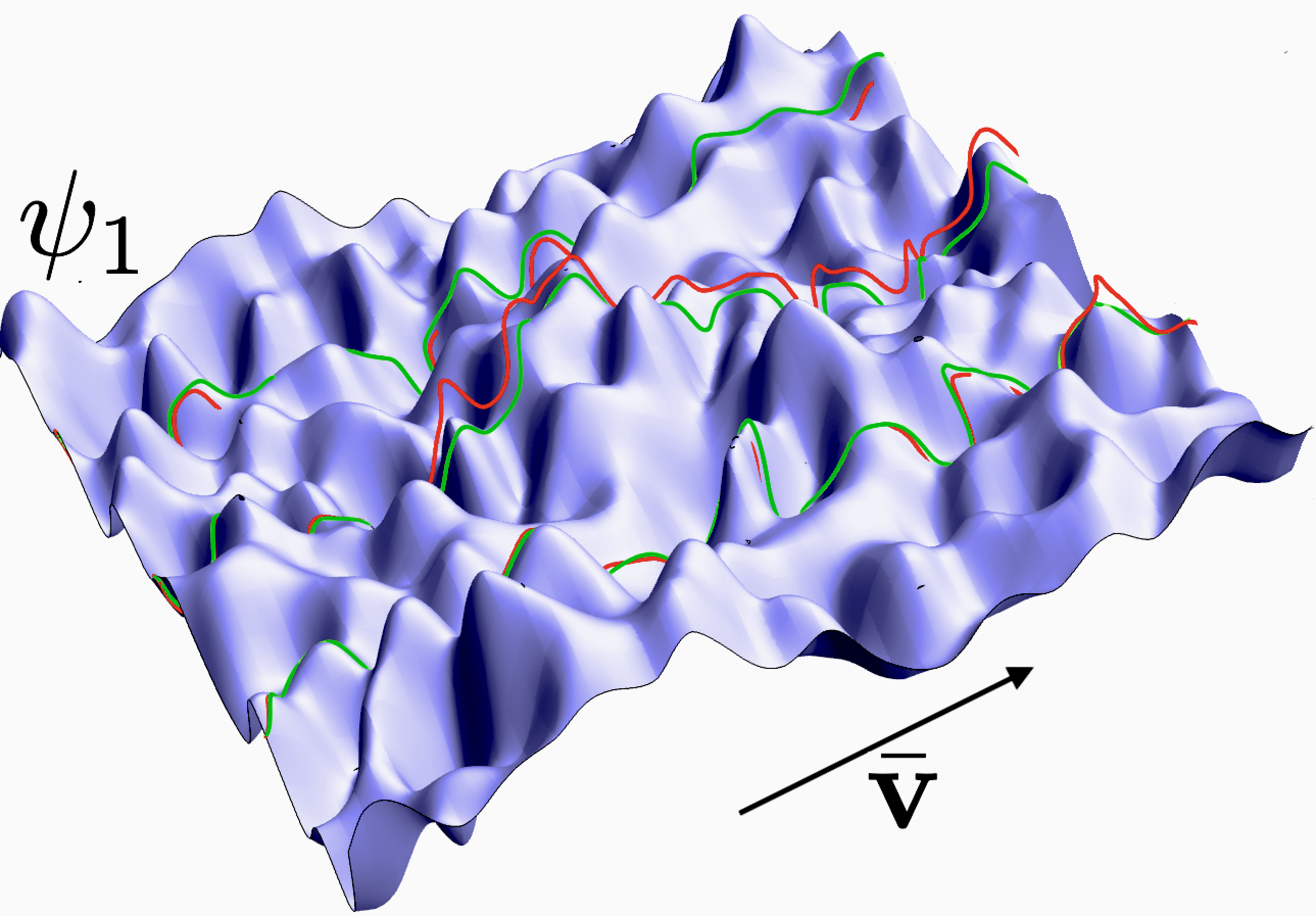}&
\includegraphics[width=0.43\columnwidth]{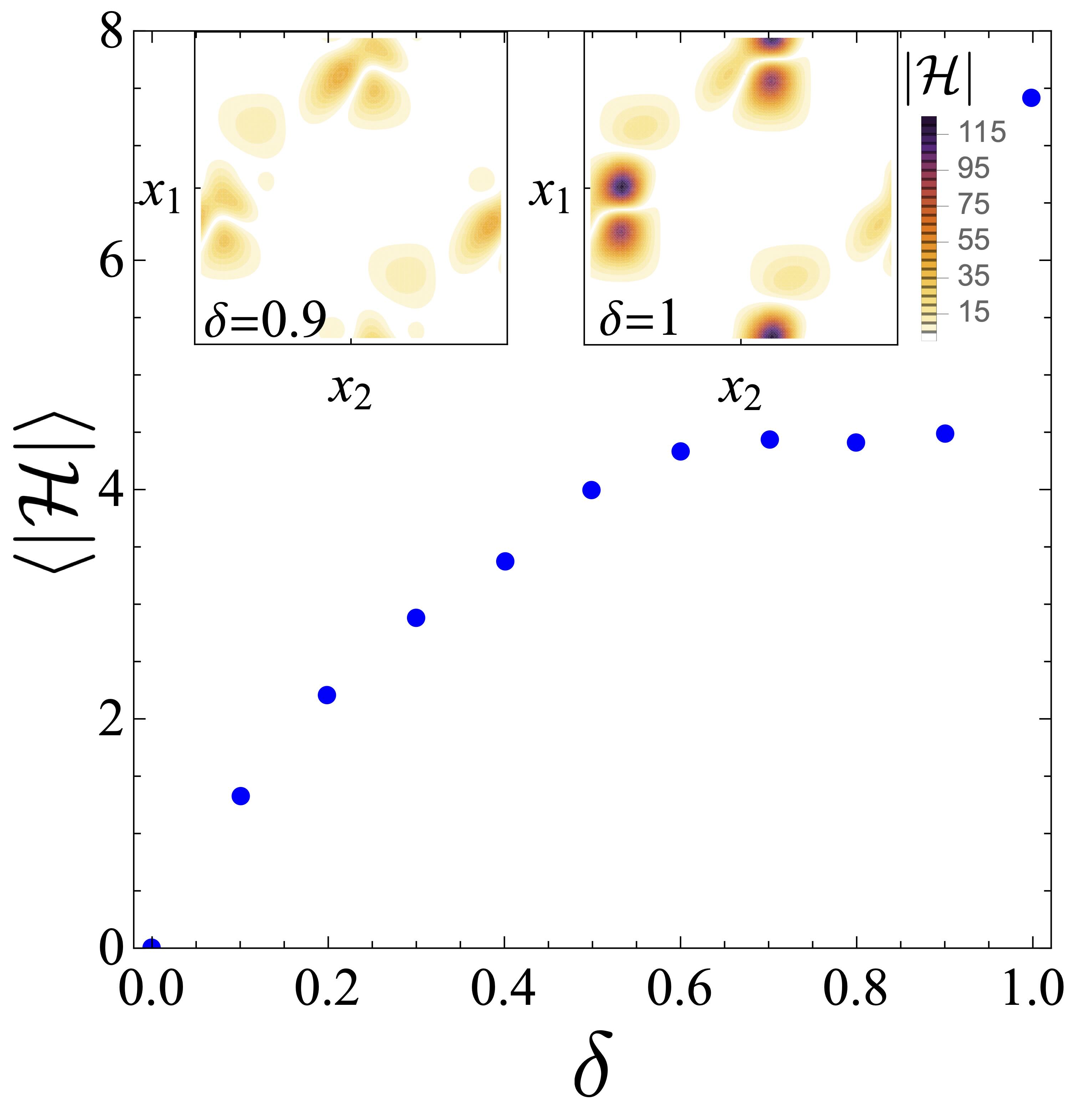}\\
(a) & (b)\\
\includegraphics[width=0.47\columnwidth]{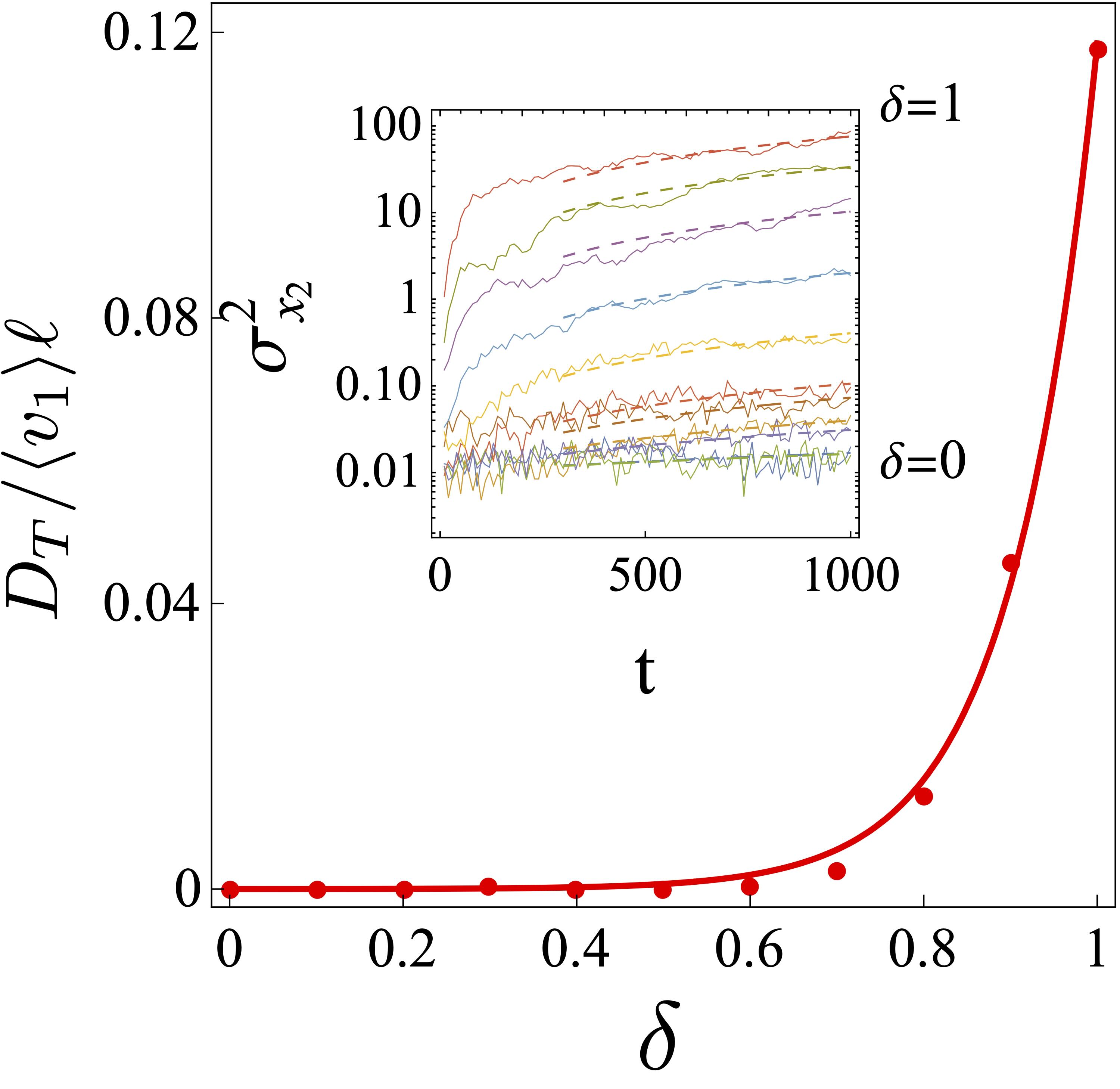}&
\includegraphics[width=0.45\columnwidth]{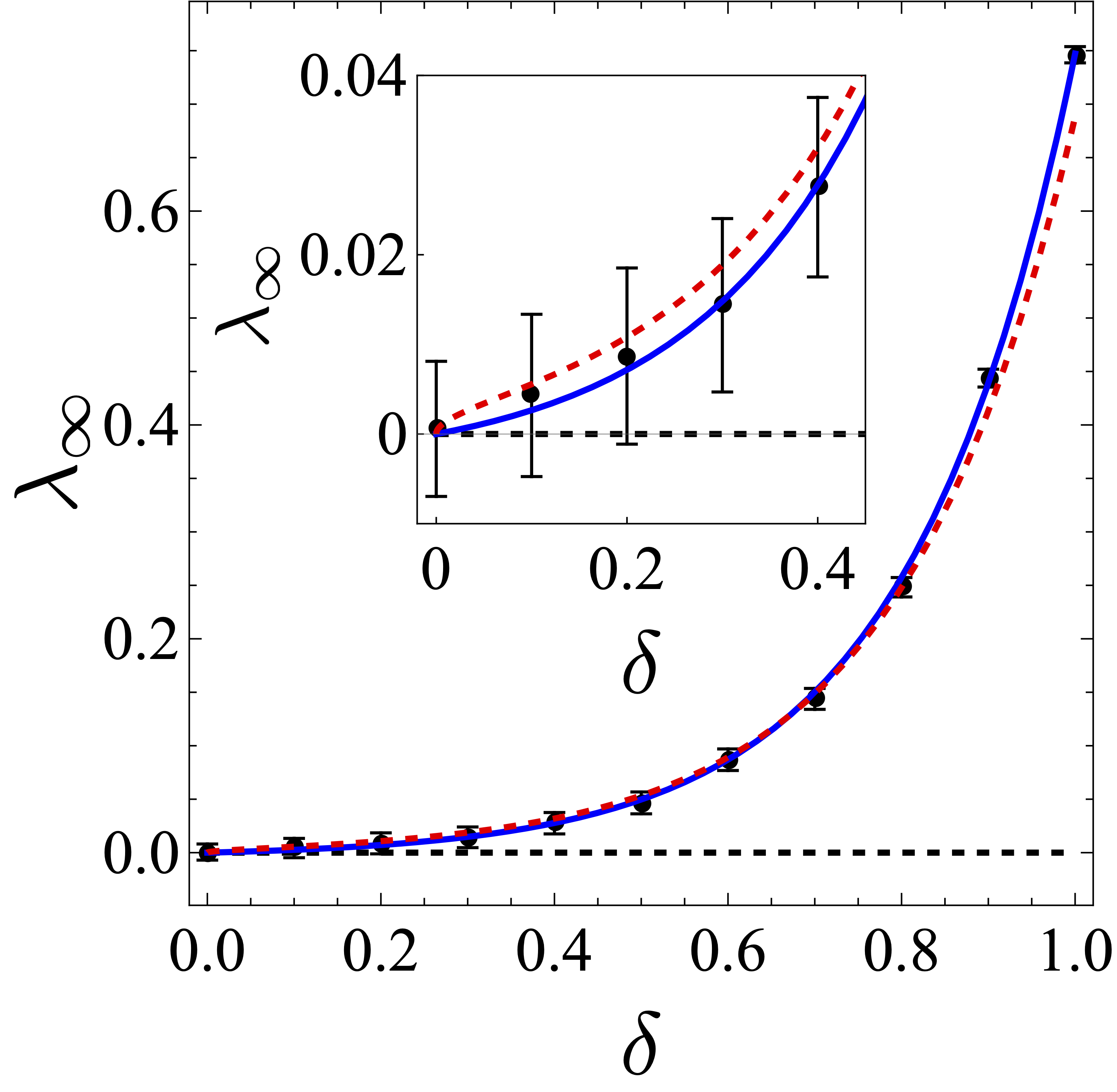}\\
(c) & (d)
\end{tabular}
\end{centering}
\caption{
(a) Perturbation of $\delta=0.1$ streamlines (red) from $\delta=0$ (zero helicity) streamlines (green) and associated $\psi_1$ streamsurface (blue) for the conductivity field given in (\ref{eqn:perturb}). Similar perturbation of $\delta=0.1$ streamlines away from the $\psi_2$ streamsurfaces (not shown) also occurs.
(b) Growth of mean absolute helicity $\langle|\mathcal{H}|\rangle$ with $\delta$, inset shows $|\mathcal{H}|$ fields for $\delta =0.9, 1$ (Adapted from \citep{Lester:2024aa}).
(c) Growth of transverse dispersivity $D_T/\langle v_1\rangle \ell$ with $\delta$ from simulations (red points) and fitted exponential (\ref{eqn:DT_delta}) (red curve). Inset shows temporal evolution of transverse variance. 
(d) Growth of Lyapunov exponent $\lambda_\infty$ with perturbation parameter $\delta$ from simulations (black points) and fitted exponential (\ref{eqn:lambda_delta}) (blue curve). (red dotted curve) dimensionless Lyapunov exponent $\lambda_\infty$ predicted from fitted exponential in (b) and (\ref{eqn:model}). (c), (d) Adapted from \citep{Lester:2024aa}.
}
\label{fig:delta}
\end{figure}

To solve the divergence-free condition $\nabla\cdot\mathbf{v}=0$ over $\Omega$, $\phi$ is decomposed into a mean and fluctuation as 
\begin{equation}
\phi(\mathbf{x})=\bar\phi(\mathbf{x})+\tilde\phi(\mathbf{x}),\label{eqn:phidecomp}
\end{equation}
where $\bar\phi=-x_1$. From (\ref{eqn:Darcy}), $\tilde{\phi}(\mathbf{x})$ is governed by 
\begin{equation}
\nabla\cdot(\mathbf{K}\cdot\nabla\tilde{\phi}(\mathbf{x}))-
\nabla\cdot(\mathbf{K}\cdot\hat{\mathbf{e}}_1)=0.\label{eqn:potfluc}
\end{equation}
Methods to solve (\ref{eqn:potfluc}) and perform streamline tracking are detailed in Appendix~\ref{app:numerics}, and typical potential fields and streamlines are shown in Figure~\ref{fig:fields}. For each value of $\delta$, a realization of $\mathbf{K}(\mathbf{x})$ is generated, (\ref{eqn:potfluc}) is solved and $10^3$ 3D streamlines are computed for distance $10^4 \ell$, along with the transverse dispersivity $D_T$, Protean velocity gradient tensor $\boldsymbol\epsilon^\prime$ and Lyapunov exponent $\lambda_\infty=\langle\epsilon_{22}^\prime\rangle$, as described in Appendix~\ref{app:1Dstretch}.

Typical streamfunctions $\psi_1(\mathbf{x})$, $\psi_2(\mathbf{x})$ for the helicity-free flow $\delta=0$ given by (\ref{eqn:streams}) (see Appendix~\ref{app:numerics} for solution details) are shown in Figure~\ref{fig:delta}a. Streamlines (red) are confined to the intersections of the level sets of $\psi_1$ and $\psi_2$, and their global behaviour is shown in Figure~\ref{fig:fields}c. Despite their significant tortuosity, these confined streamlines do not exhibit persistent dispersion or non-trivial braiding. Conversely, for $\delta>0$ the associated streamlines (green) in Figure~\ref{fig:delta}a are unconfined and so exhibit streamline braiding and transverse dispersion, see also Figure~\ref{fig:fields}d for $\delta=1$.

Figure~\ref{fig:delta}b shows that the mean helicity magnitude $\langle|\mathcal{H}|\rangle$ increases from $\mathcal{H}(\mathbf{x})=0$ everywhere for $\delta=0$ to a plateau $\langle|\mathcal{H}|\rangle\approx 4.74$ at $\delta=0.9$, then suddenly increases to $\langle|\mathcal{H}|\rangle\approx 7.16$ at $\delta=1$. This sharp increase as $\delta\uparrow 1$ is attributed to the loss of correlation between the $K_{11}$ and $K_{22}=K_{33}$ fields. The linear growth of $\langle|\mathcal{H}|\rangle$ for $\delta\ll 1$ is explained by decomposing the potential field $\phi$ as $\phi(\mathbf{x})=\phi_0(\mathbf{x})+\delta\,\phi_\delta(\mathbf{x})$. To leading order  $\mathbf{v}(\mathbf{x})$ is then
\begin{equation}
\begin{split}
\mathbf{v}(\mathbf{x})&=\mathbf{v}_0(\mathbf{x})+\delta \mathbf{v}_\delta(\mathbf{x}),\\
&=-k_0\nabla\phi_0-\delta[k_0\nabla\phi_\delta-k_\delta \hat{\mathbf{e}}_1(\hat{\mathbf{e}}_1\cdot\nabla\phi_0)]+\mathcal{O}(\delta^2)
,\end{split}
\end{equation}
where $\mathbf{v}_0(\mathbf{x})$ is helicity-free and the helical perturbation $\mathbf{v}_\delta(\mathbf{x})$ is generated by the difference $k_0-k_\delta$
\begin{equation}
\begin{split}
\mathcal{H}(\mathbf{x})&=\delta[k_0\nabla\phi\cdot(\hat{\mathbf{e}}_1\times\nabla(k_\delta-k_0)(\nabla\phi\cdot\hat{\mathbf{e}}_{1}))\\
&+(k_\delta-k_0)(\nabla\phi\cdot\hat{\mathbf{e}}_{1}\otimes \hat{\mathbf{e}}_1\cdot(\nabla k_0\times\nabla\phi))]+\mathcal{O}(\delta^2)
.\label{eqn:hel_peturb}
\end{split}
\end{equation}

Figure~\ref{fig:delta}c shows $D_T$ increases exponentially with $\delta$ as 
\begin{equation}
\frac{D_T}{\langle v_1\rangle \ell}=\frac{1}{Pe_T}\approx 4.343\times 10^{-6}(e^{10.214\,\delta}-1),\label{eqn:DT_delta}
\end{equation}
and (\ref{eqn:streams}) recovers $D_T=0$ for $\delta=0$~\citep{Lester:2023aa}, indicating finite transverse dispersion arises for weak perturbations away from heterogeneous isotropic Darcy flow. Figure~\ref{fig:delta}d shows that $\lambda_\infty$ also increases exponentially with $\delta$ as
\begin{equation}
\lambda_\infty\approx 0.00381(e^{5.283\,\delta}-1),\label{eqn:lambda_delta}
\end{equation}
 and is also non-zero for small $\delta>0$, indicating that chaotic advection occurs for weak perturbations away from isotropic Darcy flow. For $\delta=0$ the streamfunction formulation (\ref{eqn:streams}) enforces zero Lyapunov exponent~\citep{Lester:2022aa}. Figure~\ref{fig:delta}d also shows that insertion of the fitted exponential (\ref{eqn:lambda_delta}) for $D_T$  into (\ref{eqn:model}) (assuming $\Delta_L=\ell$) yields excellent agreement with the measured Lyapunov exponent (\ref{eqn:DT_delta}). This is unexpected as the domain $\Omega$ is finite and so does not belong to a universality class~\citep{Odor:2004aa}; however, these results indicate that the relationship (\ref{eqn:model}) can also hold away from the limit $N_p=L/\ell\rightarrow\infty$. These results establish that chaotic advection is inherent to heterogeneous Darcy flow, even in the limit of weakly anisotropic media.

\subsection{Onset of Chaotic Advection with Medium Heterogeneity}\label{sec:heterogeneity}

To examine the impact of medium heterogeneity upon anisotropic Darcy flow, we consider flow generated by the anisotropic diagonal conductivity tensor $\mathbf{K}(\mathbf{x})$ in (\ref{eqn:aniso}),
where the scalar conductivity fields $K_{ii}$ are generated in the same manner as $k_0$, $k_\delta$ but with $N_i=4$. The heterogeneity of these fields $K_{ii}$ is varied from weakly ($\sigma^2_{\ln K}<1$) to strongly ($\sigma^2_{\ln K}>1$) heterogeneous over the range of log-variances $\sigma^2_{\ln K}=$$(2^{-10},2^{-8},\dots, 2^{-2})$ $\cup$ $(1/2,1,2,3,4)$. For each value of $\sigma^2_{\ln K}$, a realisation of $\mathbf{K}(\mathbf{x})$ is generated and (\ref{eqn:potfluc}) is solved, and streamlines, $D_T$ and $\lambda_\infty$ are computed in the same manner as described in \S\ref{sec:anisotropy}.

\begin{figure}
\begin{centering}
\begin{tabular}{c c}
\includegraphics[width=0.48\columnwidth]{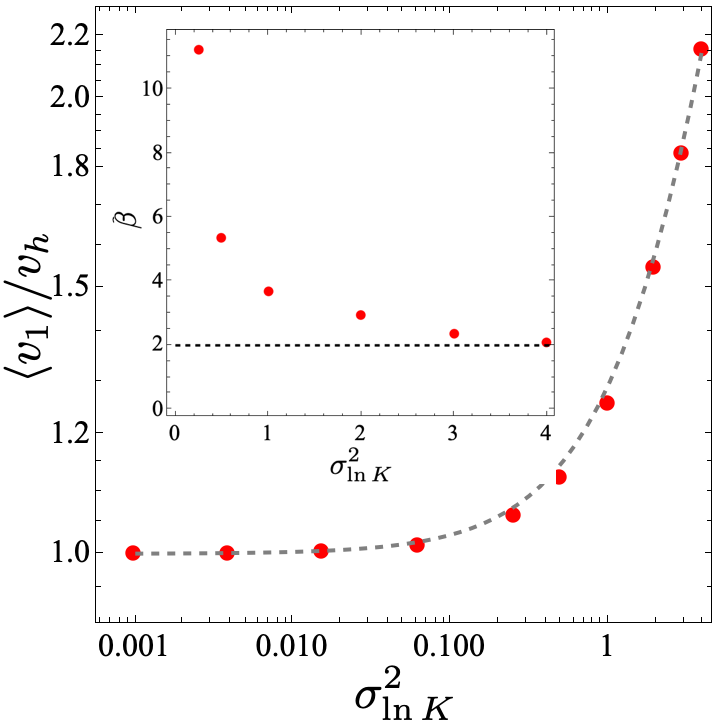}&
\includegraphics[width=0.47\columnwidth]{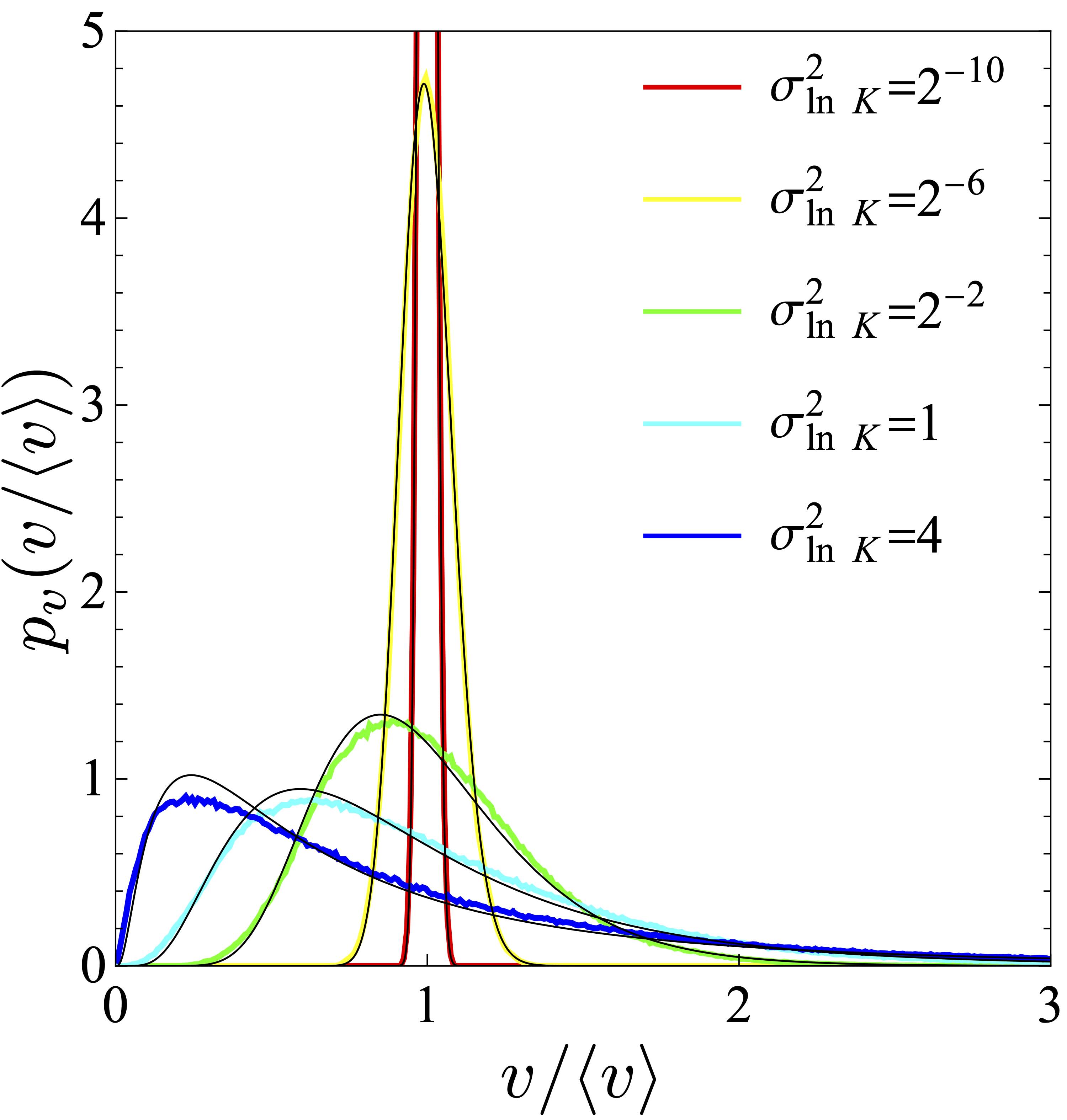}\\
(a) & (b)\\
\includegraphics[width=0.48\columnwidth]{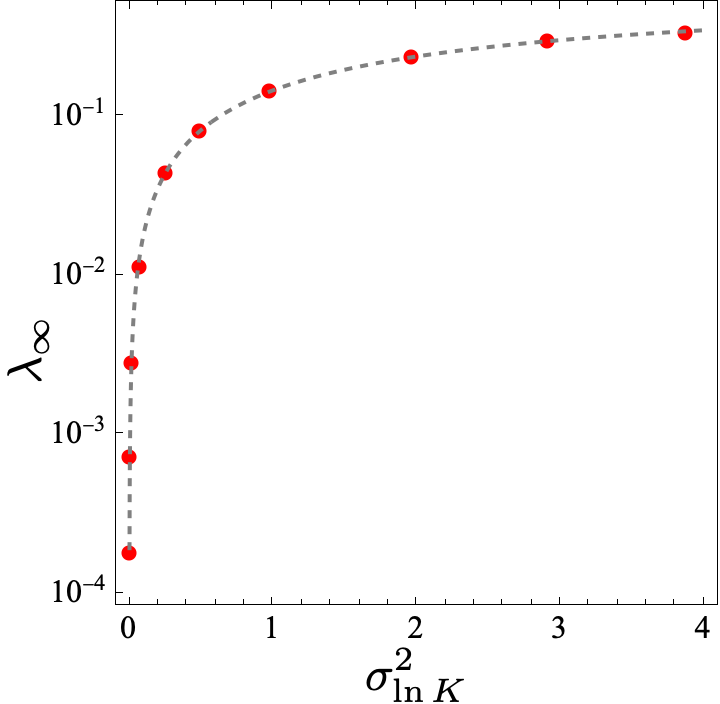}&
\includegraphics[width=0.49\columnwidth]{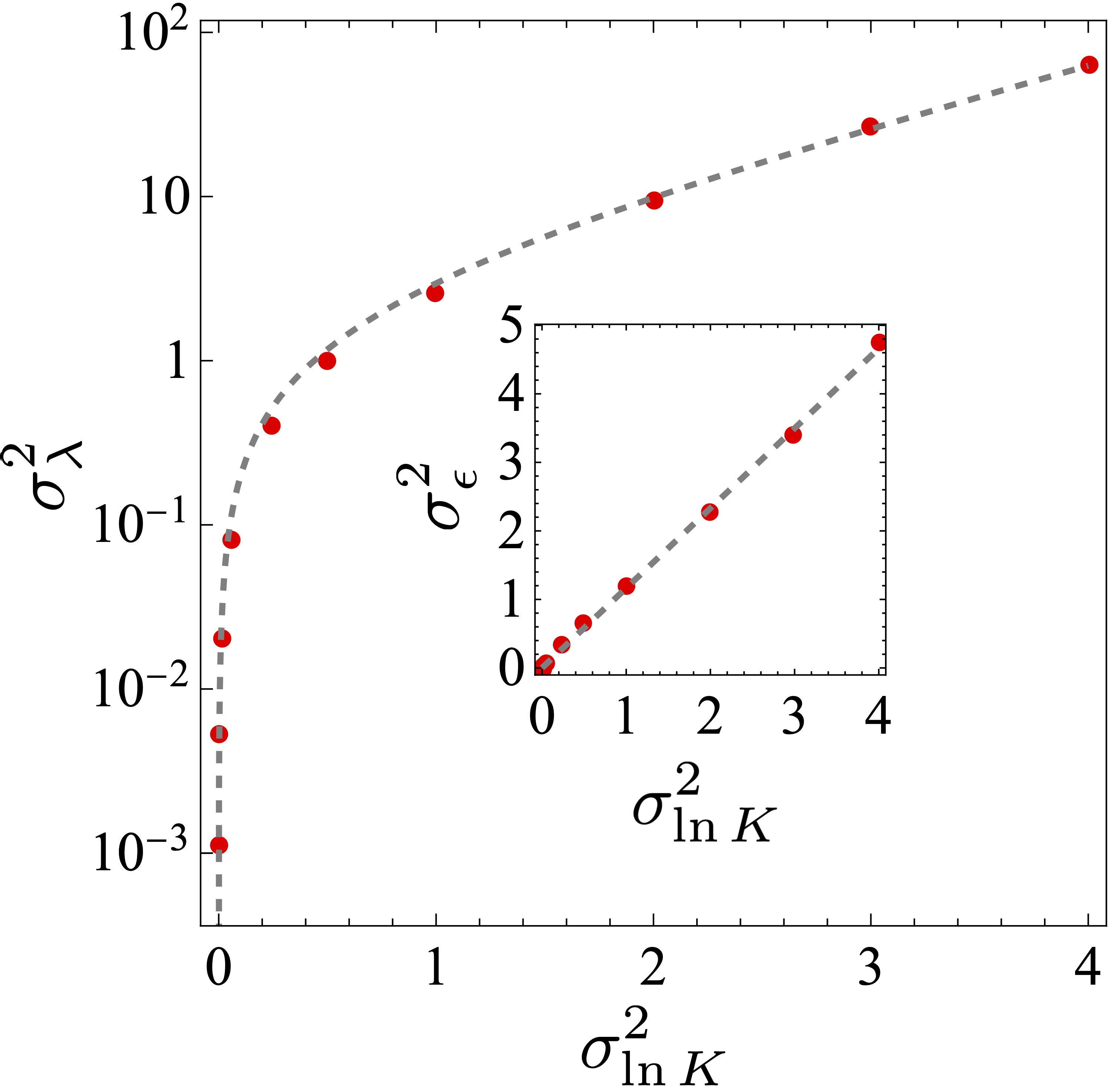}\\
(c) & (d)
\end{tabular}
\end{centering}
\caption{(a) Variation of normalised mean longitudinal velocity $\langle v_1\rangle/v_h$ (red points) with log-variance $\sigma^2_{\ln K}$ and linear fit (\ref{eqn:vmean}), (grey dashed line) in anisotropic Darcy flow. (Inset) Variation of small velocity scaling index $\beta$ with $\sigma^2_{\ln K}$. (b) PDFs of normalised velocity magnitude $p_v(v/\langle v\rangle)$ as a function of log-variance $\sigma^2_{\ln K}$ and fitted log-normal distribution (black lines). (c) Variation of dimensionless Lyapunov exponent $\lambda_\infty$ (red dots) with log-variance $\sigma^2_{\ln K}$ and nonlinear fit (\ref{eqn:lyapunov_fit}), (grey dashed line). (d) Variation of dimensionless stretching variance $\sigma^2_\lambda$ with log-variance $\sigma^2_{\ln K}$ and nonlinear fit (\ref{eqn:lnrhovar}), (grey dashed line). (Inset) Variation of Protean velocity gradient variance $\sigma^2_\epsilon$ with log-variance $\sigma^2_{\ln K}$ and linear fit (\ref{eqn:velgradvar}), (grey dashed line).}
\label{fig:velpdfs}
\end{figure}

Figure~\ref{fig:velpdfs}a shows that the scaled mean longitudinal velocity $\langle v_1\rangle/v_h$ (where $v_h$ is the velocity for a homogeneous medium) increases linearly with log-variance $\sigma^2_{\ln K}$ as expected for small and moderate values of $\sigma^2_{\ln k}$ \citep{renard1997calculating}
\begin{equation}
\frac{\langle v_1\rangle}{v_h}= 1+\alpha_1\,\sigma^2_{\ln K},\label{eqn:vmean}
\end{equation}
with $\alpha_1\approx 0.29178$ and coefficient of determination $R^2=0.999$. Figure~\ref{fig:velpdfs}b shows that the Eulerian velocity PDF $p_v(v)$ roughly follows a log-normal distribution for all but strongly heterogeneous flows,
and converges toward a delta function in the homogeneous limit $\sigma^2_{\ln K}\rightarrow 0$. For $\sigma^2_{\ln K}\geqslant 1$, the velocity PDF scales as a power law in the small velocity limit as $p_v(v)\propto v^{\beta-1}$, and the inset in Figure~\ref{fig:velpdfs}a shows that the index $\beta>2$ and approaches $\beta\downarrow 2$ in the strongly heterogeneous regime, leading to normal transport for all $\sigma^2_{\ln K}\geqslant 1$. The persistence of normal transport in the strongly heterogeneous regime is attributed to the low probability of all three $K_{ii}$ in (\ref{eqn:aniso}) being simultaneously small. We also note that different random models such as Gamma-distributed conductivity fields have a greater propensity to generate anomalous transport~\citep{Hakoun:2019aa}. 

Figure~\ref{fig:velpdfs}c shows that $\lambda_\infty$ converges to a constant value with increasing log-variance $\sigma^2_{\ln K}$, and is well fitted by the simple nonlinear function (\ref{eqn:lyapunov_fit}) as described below. As shown in the inset of Figure~\ref{fig:velpdfs}d, the velocity gradient variance $\sigma^2_\epsilon$ (from data presented in \S\ref{sec:velgrad}) grows linearly with conductivity log-variance as
\begin{equation}
\sigma^2_\epsilon\approx\alpha_2\sigma^2_{\ln K},\label{eqn:velgradvar}
\end{equation}
where $\alpha_2=1.1656$ with $R^2=0.999$, and $\sigma^2_\lambda$ grows exponentially with $\sigma^2_{\ln K}$ as
\begin{equation}
\sigma^2_\lambda\approx\alpha_3\left(e^{\alpha_4\sigma^2_{\ln K}}-1\right),\label{eqn:lnrhovar}
\end{equation}
where $\alpha_3=2.2429$ and $\alpha_4=0.8415$ with $R^2=0.999$. These results show that although the Lyapunov exponent $\lambda_\infty$ converges toward the upper bound $\ln 2$ with increasing medium heterogeneity, the corresponding variance $\sigma^2_\lambda$ increases due to growth of the second moment $\langle \tau^2\rangle$. For the case of anomalous transport ($1<\beta<2$), this variance grows super-linearly as $\sigma^2_{\ln\rho}\sim t^{3-\beta}$~\citet{Lester:2018aa}. 

\begin{figure}
\begin{centering}
\begin{tabular}{c c}
\includegraphics[width=0.48\columnwidth]{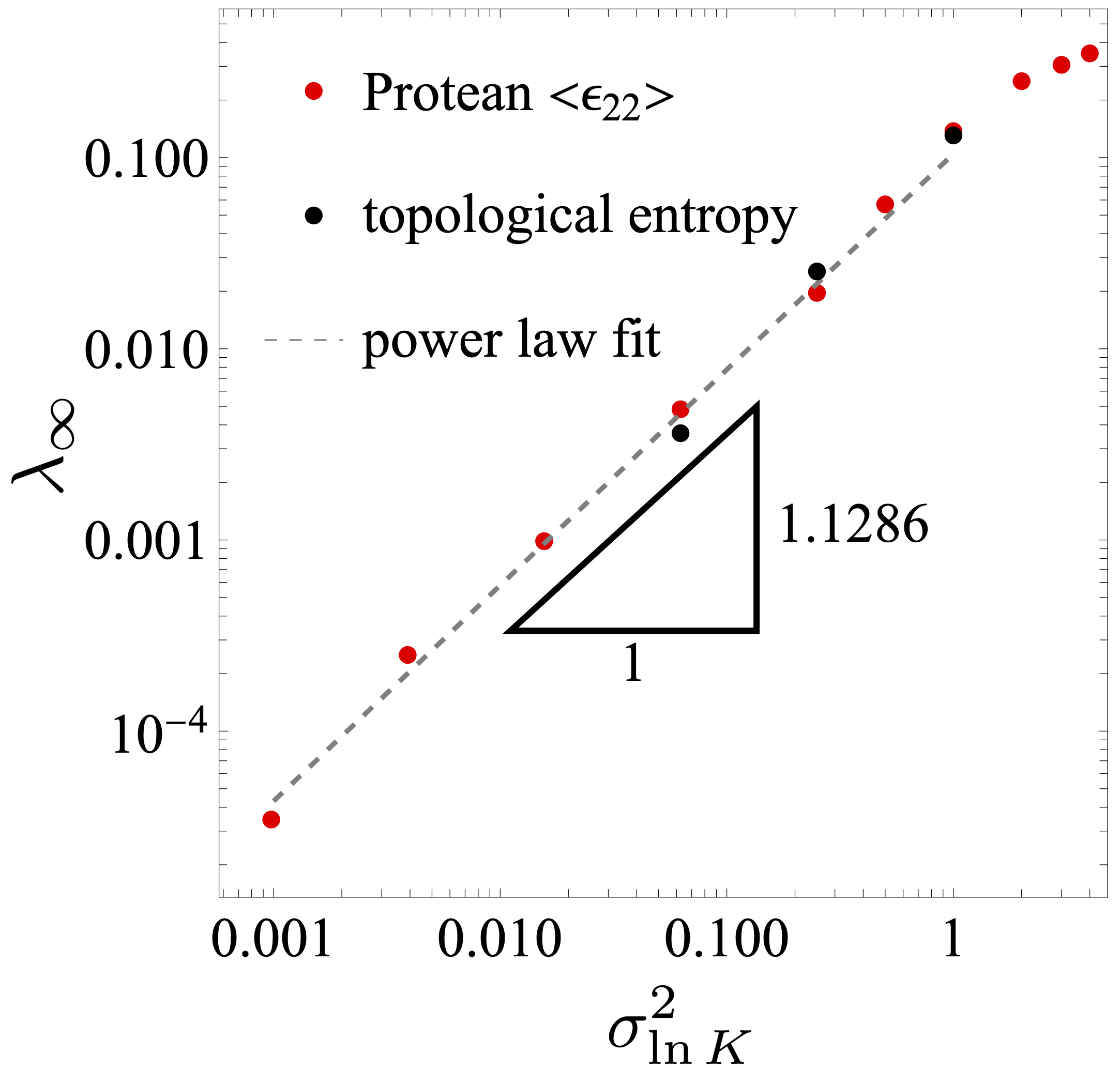}&
\includegraphics[width=0.46\columnwidth]{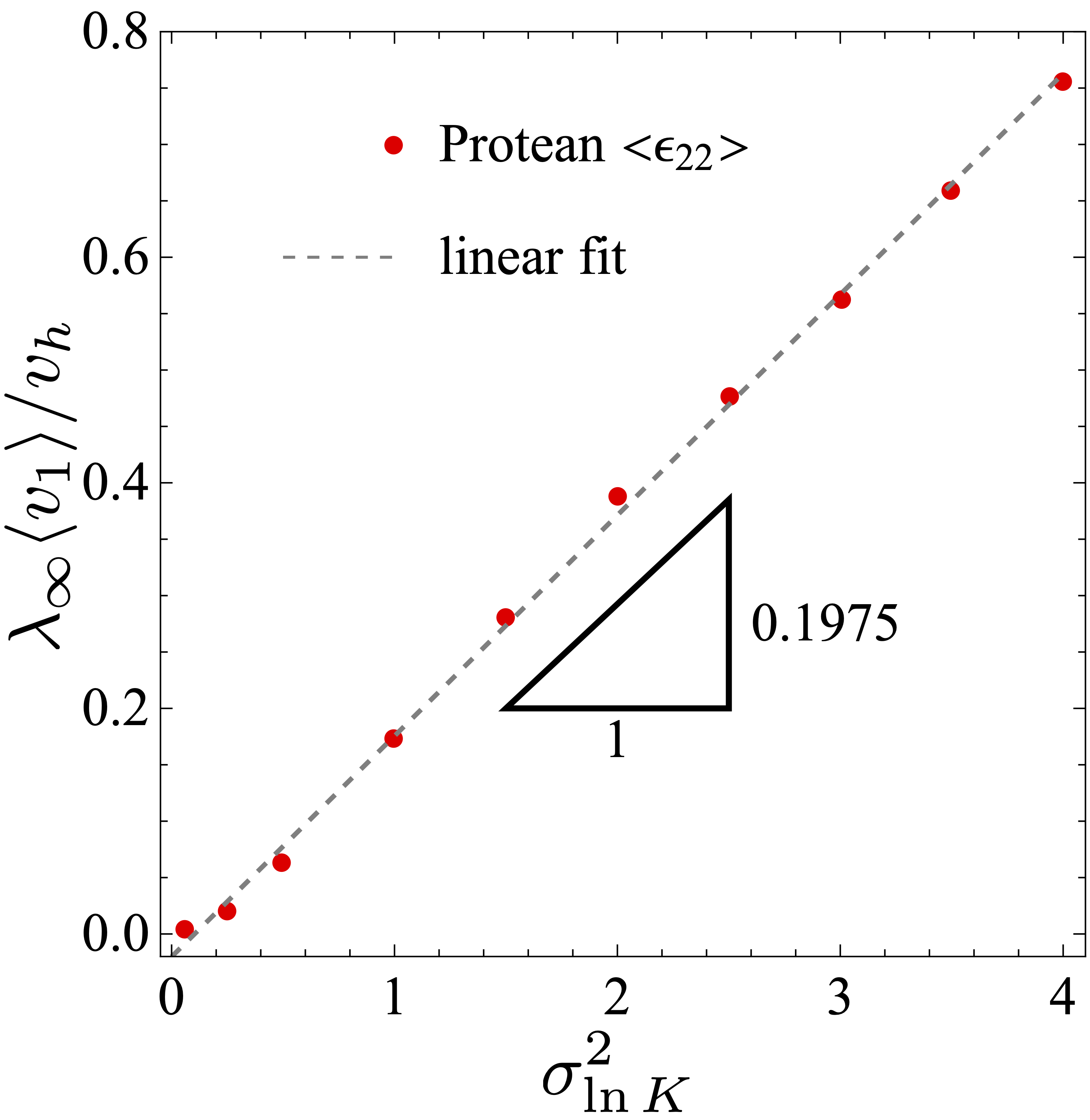}\\
 (a) & (b)\\
 \includegraphics[width=0.48\columnwidth]{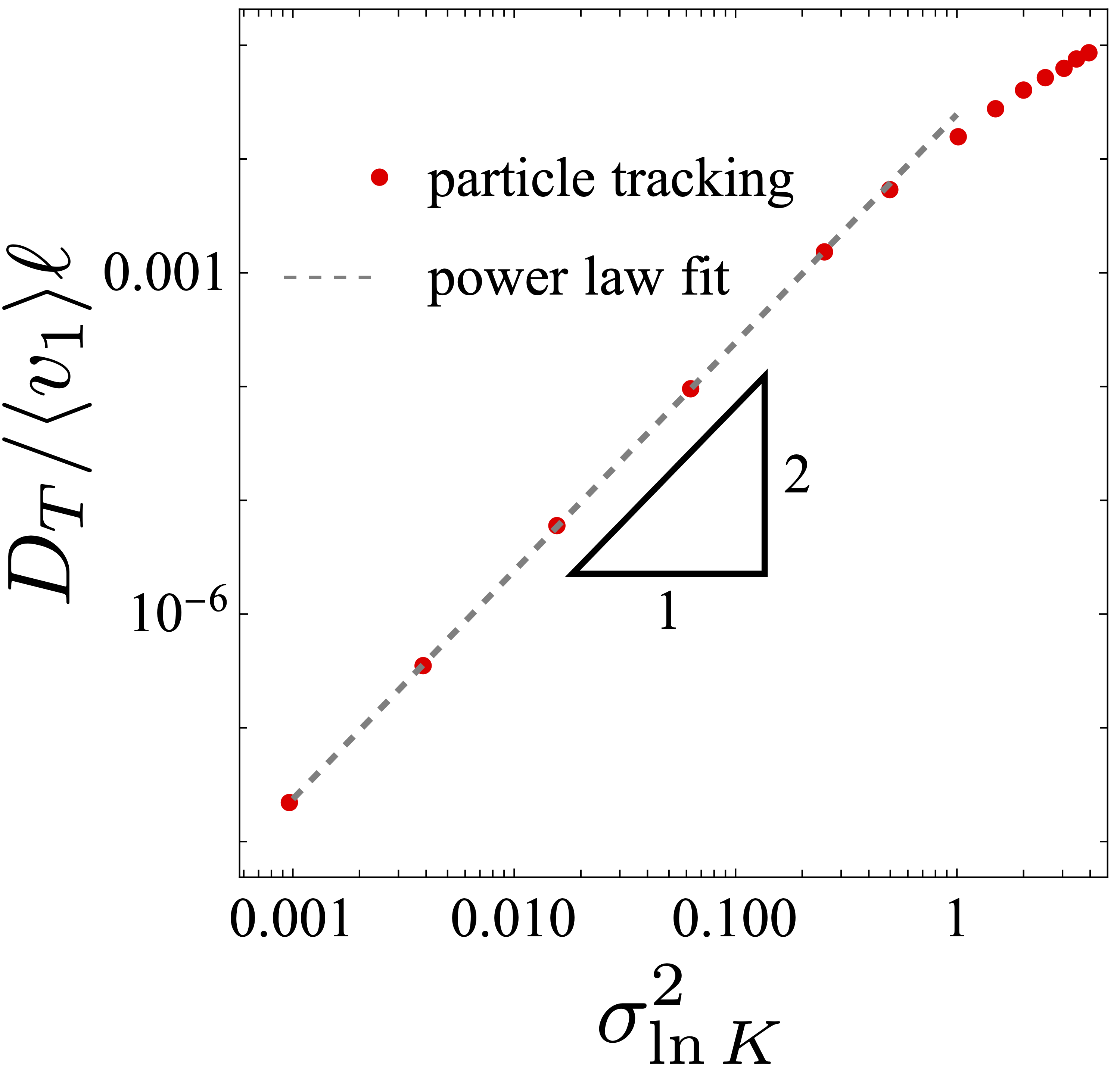}&
\includegraphics[width=0.465\columnwidth]{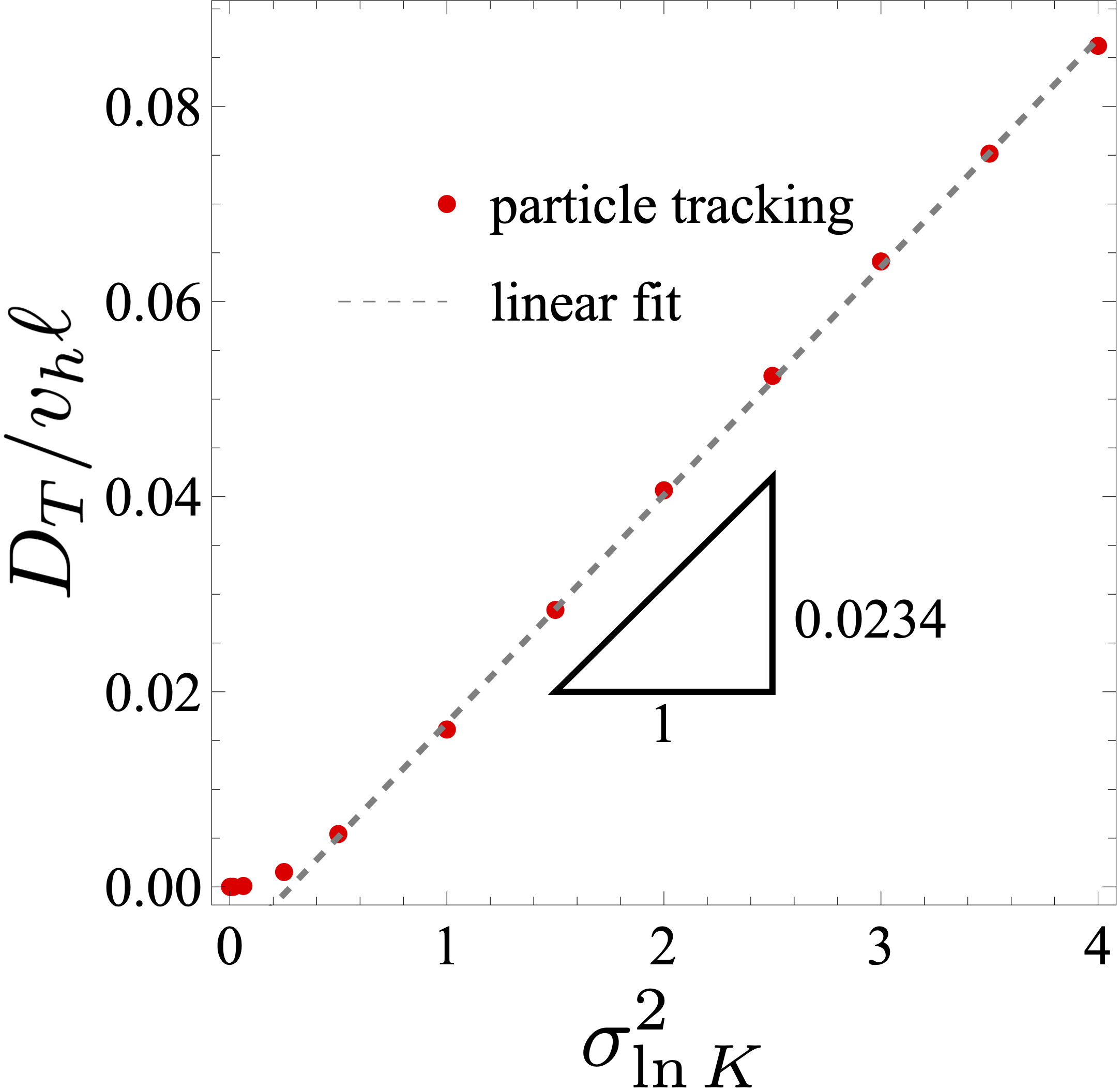}\\
 (c) & (d)
\end{tabular}
\end{centering}
\caption{(top) Lyapunov exponents and (bottom) dispersion coefficients in (a, c) weakly ($\sigma^2_{\ln K}\leqslant 1$) and (b,d) strongly ($\sigma^2_{\ln K}\geqslant 1$) heterogeneous porous media with anisotropic conductivity given by (\ref{eqn:aniso}). Red points in (a, b) indicate Lyapunov exponents computed from the Protean frame~\citep{Lester:2018aa}, black points indicate topological entropy computed using E-tec~\citep{Roberts:2019aa}. Dashed lines in (a, c) and (b, d) respectively indicate power law (\ref{eqn:lambda_power}), (\ref{eqn:DT_power}), and linear (\ref{eqn:lambda_rescale}), (\ref{eqn:DT_linear}), fits to the Lyapunov exponents and dispersion coefficients in the weakly and strongly heterogeneous regimes.}
\label{fig:lyapunov}
\end{figure}

In addition to direct computation of $\lambda_\infty$, the topological braid entropy $h_\text{braid}$ of the streamlines is computed directly via the E-tec routine~\citep{Roberts:2019aa}, which is limited to moderately heterogeneous ($0.1\leqslant \sigma^2_{\ln K}\leqslant 2$) systems due to lack of convergence for $\sigma^2_{\ln K}<0.1$ and flow reversal for $\sigma^2_{\ln K}>1$. In principle, streamlines for the strongly heterogeneous cases could be placed in the intrinsic coordinates $\boldsymbol\xi=(\chi_1,\chi_2,\phi)$, however, determination of $\chi_1$, $\chi_2$ is difficult as the non-zero helicity density $\mathcal{H}(\mathbf{x})\neq 0$ means that mutual Lie derivatives of the flow do not vanish and so there does not exist an intrinsic \emph{holonomic basis}~\citep{Schutz:1980aa} for $(\chi_1,\chi_2)$. Figure~\ref{fig:lyapunov}a shows that $\lambda_\infty$ and $h_\text{braid}$ agree to within 5\%, verifying (\ref{eqn:entropy_mean}).


In the weakly heterogeneous regime ($\sigma^2_{\ln K}\leqslant 1$), $\lambda_\infty$ varies almost linearly with log-variance as 
\begin{equation}
\lambda_\infty\approx (\sigma^2_{\ln K})^{\alpha_5},\label{eqn:lambda_power}
\end{equation}
with $\alpha_5=1.129$ and $R^2=0.995$, suggesting that chaotic advection arises in weakly heterogeneous anisotropic conductivity fields. Figure~\ref{fig:lyapunov}b shows that in the strongly heterogeneous regime, $\lambda_\infty$ scales linearly with $\sigma^2_{\ln K}$ as
\begin{equation}
\lambda_\infty\frac{\langle v_1\rangle}{v_h}\approx \alpha_6\,\sigma^2_{\ln K},\label{eqn:lambda_rescale}
\end{equation}
with $\alpha_6\approx 0.198$ and $R^2=0.999$. As per Figure~\ref{fig:velpdfs}c, the behavior of $\lambda_\infty$ as a function of $\sigma^2_{\ln K}$ can be fitted by 
\begin{equation}
\lambda_\infty=\frac{\alpha_1 \sigma^2_{\ln K}}{1+\alpha_6\sigma^2_{\ln K}},\label{eqn:lyapunov_fit}
\end{equation}
which implies that $\lambda_\infty$ converges in the strongly heterogeneous limit
\begin{equation}
\lim_{\sigma^2_{\ln K}\rightarrow\infty}\lambda_\infty\approx \frac{\alpha_1}{\alpha_6}= 0.677.\label{eqn:converge}
\end{equation}
This value is close to the upper bound $h=\ln 2\approx 0.693$ for steady 3D flow~\citep{Dinh:2008aa}, indicating strong chaotic mixing arises in strongly heterogeneous Darcy flow. 

Convergence of $\lambda_\infty$ in (\ref{eqn:converge}) is understood by considering the limit $\sigma^2_{\ln K}\rightarrow\infty$ and partitioning $\mathbf{K}$ over $\Omega$ into infinitely permeable $\Omega_\infty$ ($\text{tr}\,\mathbf{K}(\mathbf{x})\rightarrow\infty$) and an impermeable $\Omega_0$ ($\text{tr}\,\mathbf{K}(\mathbf{x})\rightarrow 0$) subdomains. As $\Omega$ is the 3-torus $\mathbb{T}^3$ with topological genus $g=3$, the boundary $\partial\Omega_{\infty 0}$ between the subdomains is also topologically complex and thus admits saddle-type stagnation points $\mathbf{x}_0$ on $\partial\Omega_{\infty 0}$ as per the Poincar\'{e}-Hopf theorem. These saddle points generate chaotic advection via the same mechanism as for pore-scale Stokes flow~\citep{Lester:2013aa}. Due to the linearity of Darcy flow, the rescaled fluid velocity field $\mathbf{v}/\langle v_1\rangle$ and associated saddle points and Lyapunov exponent are all invariant in the limit $\sigma^2_{\ln K}\rightarrow\infty$.

Figure~\ref{fig:lyapunov}c shows that the transverse dispersivity scales nonlinearly in the weakly heterogeneous regime as
\begin{equation}
\frac{D_T}{\langle v_1\rangle\ell}\sim (\sigma^2_{\ln K})^{2.01},\label{eqn:DT_power}
\end{equation}
and linearly in the strongly heterogeneous regime as
\begin{equation}
\frac{D_T}{v_h\ell}\approx 0.0234\,\sigma^2_{\ln K},\label{eqn:DT_linear}
\end{equation}
with $R^2=1.000$ in both cases. In the weakly heterogeneous regime, the scalings of $D_T$ (\ref{eqn:DT_power}) and $\lambda_\infty$ (\ref{eqn:lambda_power}) with $\sigma^2_{\ln K}$ agree with the relationship (\ref{eqn:model}) linking dispersion and chaotic advection in random flows. However, as expected this relationship does not persist in the strongly heterogeneous regime due to the onset of flow reversal. These results establish that chaotic advection is inherent to anisotropic media, even in the limit of weakly heterogeneous conductivity fields.

\subsection{Velocity Gradient Statistics}\label{sec:velgrad}

\begin{figure}
\begin{centering}
\begin{tabular}{c c}
\includegraphics[width=0.48\columnwidth]{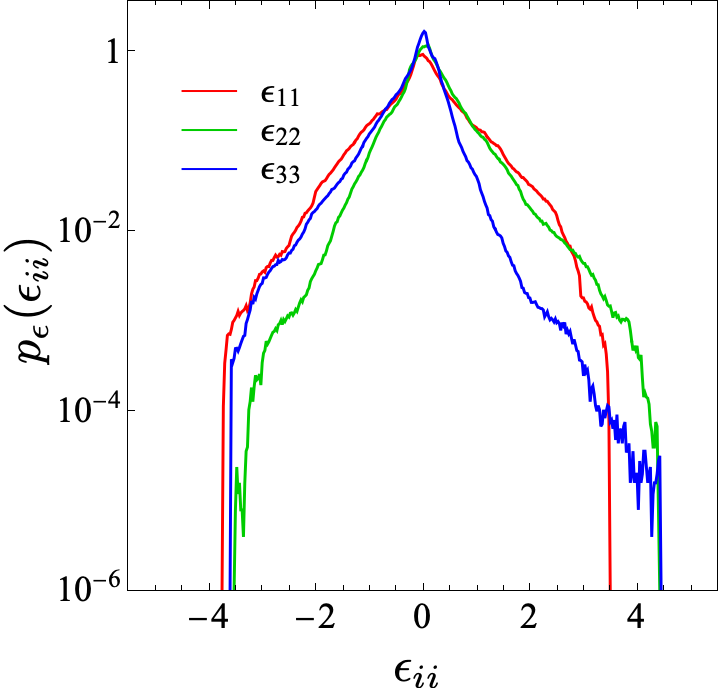}&
\includegraphics[width=0.49\columnwidth]{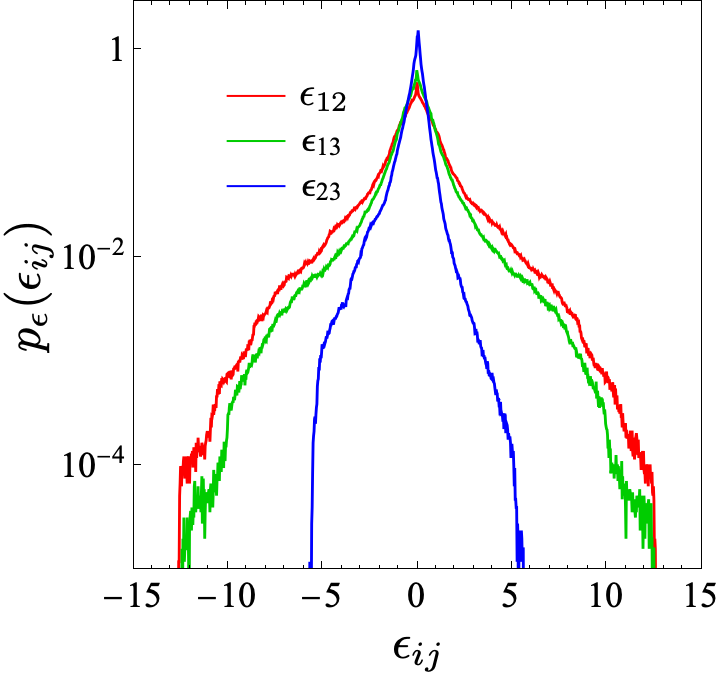}\\
(a) & (b)\\
\includegraphics[width=0.48\columnwidth]{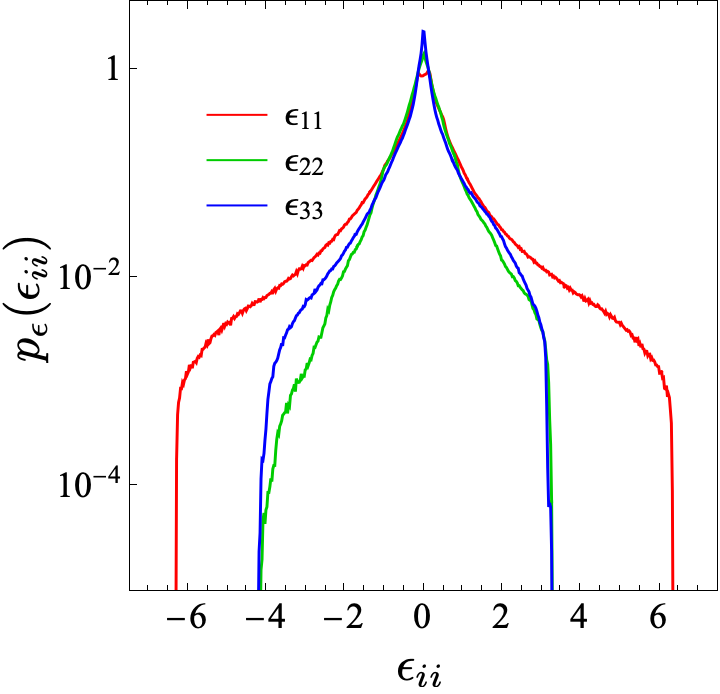}&
\includegraphics[width=0.49\columnwidth]{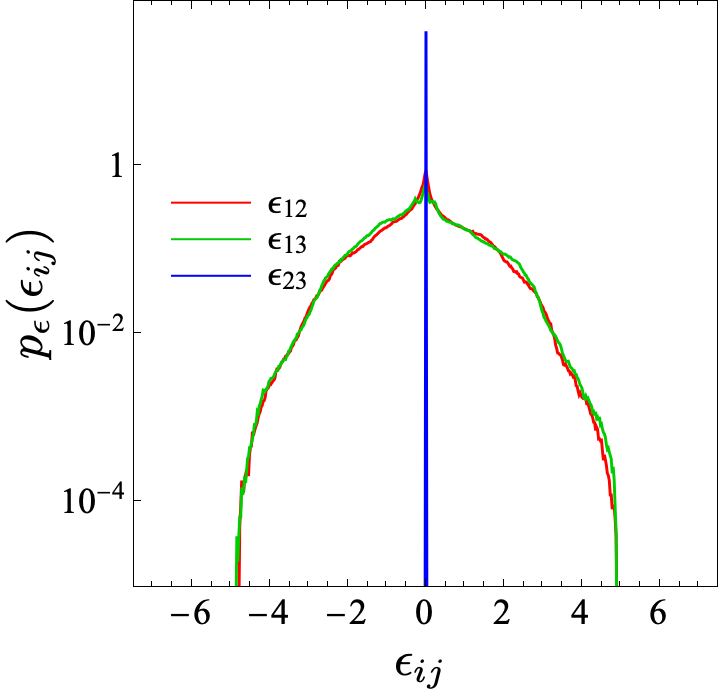}\\
(c) & (d)
\end{tabular}
\end{centering}
\caption{PDFs of (a,c) diagonal $\epsilon_{ii}$ and (b,d) off-diagonal velocity gradient components $\epsilon_{ij}$ for (top) heterogeneous anisotropic Darcy flow (\ref{eqn:perturb}) with $\sigma_{\ln K}^2=4$, $\delta=1$ and (bottom) isotropic Darcy flow  (\ref{eqn:perturb}) with $\sigma_{\ln K}^2=4$, $\delta=0$.}
\label{fig:lyapunov_pdf}
\end{figure}

In addition to the Lyapunov exponent, the Protean velocity gradient $\boldsymbol\epsilon^\prime$ provides important information regarding the deformation structure of heterogeneous Darcy flow. For all computed flows $\boldsymbol\epsilon^\prime$ is sampled along $10^3$ streamlines at fixed spatial increment $\ell$ for a distance of $10^4\ell$. Figure~\ref{fig:lyapunov_pdf}(a,c) and (b,d) respectively show the PDFs of the principal stretches $\epsilon_{ii}^\prime$ and shears $\epsilon_{ij}^\prime$ for strongly heterogeneous ($\sigma^2_{\ln K}=4$) Darcy flow for (a,b) fully anisotropic ($\delta=1$) and (c,b) isotropic ($\delta=0$) conductivity fields (\ref{eqn:perturb}). For all flows computed the divergence error $|\sum_{i=1}^3\epsilon^\prime_{ii}|<10^{-6}$ and the stream-wise mean stretch $|\langle\epsilon^\prime_{11}(t)\rangle|<10^{-4}$ is also close to zero. For all flows the distributions of both $\epsilon^\prime_{ii}$ and $\epsilon^\prime_{ij}$ for $j>i$, $i=1:3$ are broad and exhibit a sharp cutoff due to the finite nature of $\Omega$. In all cases, the standard deviation is large $\sigma_{\epsilon^\prime_{ii}}\gg |\langle\epsilon^\prime_{ii}\rangle|$, however the large number of independent observations ($10^7$) reduces the standard error to $\sigma_{\epsilon_{ii}}/\sqrt{n}\sim 10^{-3}$, yielding accurate estimates of the principle stretches.

Figure~\ref{fig:lyapunov_pdf}(a,c) shows that the PDF of $\epsilon^\prime_{11}$ is significantly broader for the isotropic case, which is attributed to confinement of streamlines to 2D streamsurfaces $\psi_1$, $\psi_2$, leading to large velocity variations. 
The off-diagonal shears $\epsilon_{ij}^\prime$ in Figure~\ref{fig:lyapunov_pdf}(b,d) are all similarly distributed except the longitudinal shears $\epsilon_{12}^\prime$, $\epsilon_{13}^\prime$ of the anisotropic flow are more broadly distributed as these streamlines are not confined to streamsurfaces and hence have higher curvature. The Protean transverse shear $\epsilon_{23}^\prime$ in the isotropic case is zero as this is related to the helicity density as $\mathcal{H}(\mathbf{x})=v^\prime_i\,\varepsilon_{ijk}\,\epsilon^\prime_{jk}=v\,\epsilon_{23}^\prime$~\citep{Lester:2018aa}. 

For isotropic flow, the Lyapunov exponent is effectively zero ($|\langle\epsilon^\prime_{ii}\rangle|<10^{-4}$), whereas for anisotropic flow it is slightly larger than the theoretical upper bound $\lambda_\infty=\ln 2$, $(\langle\epsilon^\prime_{11}\rangle,\langle\epsilon^\prime_{22}\rangle,\langle\epsilon^\prime_{33}\rangle)$=(0.0001, 0.7148, -0.7149). The broad nature of the velocity gradient PDFs leads to a relatively  large stretching variance, $\sigma^2_\lambda/\lambda_\infty\sim 10^2$ as shown in Figure~\ref{fig:velpdfs}d. This magnitude is consistent with the finding that the ensemble average  (\ref{eqn:entropy_var}) $h=\lambda_\infty+\sigma^2/2$ is not correct as this would not yield the agreement between $h_{\text{braid}}$ and $\lambda_\infty$ observed in Figure~\ref{fig:lyapunov}a. The broad distribution of all components of $\boldsymbol\epsilon^\prime$ has significant impacts upon the range of fluid processes hosted in these flows, as shall be explored further in \S\ref{sec:diffusion}.

\section{Mechanisms and Implications of Chaotic Advection}\label{sec:mechanisms_implications}

\subsection{Chaotic Advection Mechanisms}\label{sec:mechanisms}

In this section we consider the mechanisms that generate chaotic advection in steady 3D Darcy flows ranging from flows with smooth, finite $\mathbf{K}(\mathbf{x})$ to those with impermeable inclusions or stagnation points. As $v_\phi\downarrow 0$, the rescaling (\ref{eqn:vxi_prime}) breaks down~\citep{Bajer:1994aa}, leading to chaotic advection via a similar mechanism to that of pore-scale flow~\citep{Lester:2013aa}, where exponential stretching of fluid elements local to saddle-type stagnation points form stable and unstable hyperbolic manifolds and a heteroclinic tangle, the hallmark of chaos in continuous dynamical systems~\citep{Ottino:1989aa,Katok:1995aa}. In the absence of stagnation points, chaotic advection via streamline braiding arises via a different mechanism that is most clearly elucidated by consideration of spatially periodic porous media. Consider Darcy flow in a heterogeneous anisotropic hydraulic conductivity field that is $P$-periodic in the direction of the mean potential gradient $\mathbf{g}\equiv-\langle\nabla\phi\rangle/||\langle\nabla\phi\rangle||$,
\begin{equation}
\mathbf{K}(\mathbf{x})=\mathbf{K}\left(\mathbf{x}+nP\, \mathbf{g}\right),\quad n=1,2,\dots\label{eqn:periodic}
\end{equation}
From the Brouwer fixed point theorem, there must exist period-$k$ points $\mathbf{x}_k$ (with $k=1,2,\dots$) in $\Omega$, such that streamlines at $\mathbf{x}_k$ are advected downstream to $\mathbf{x}_p+kP\,\mathbf{g}$. 

Each periodic point $\mathbf{x}_k$ belongs to a periodic orbit (streamline) that can generate chaotic advection in a similar manner to stagnation points in pore-scale flow. As per Hamiltonian dynamical systems theory~\citep{Katok:1995aa}, a periodic point may be classified as a \emph{hyperbolic saddle} point if the local fluid deformation over the $k$-period involves fluid stretching and contraction that is exponential in time, the directions of which are respectively associated with hyperbolic unstable and stable manifolds emanating from $\mathbf{x}_k$. If these manifolds intersect transversely (which is inevitable in random systems), then a heteroclinic tangle results, leading to chaotic advection~\citep{Ottino:1989aa,Katok:1995aa}. For aperiodic braiding flows, hyperbolic un/stable manifolds arise via the same mechanisms but are not connected to features such as stagnation points or periodic orbits. Rather, they are detected by resolving hyperbolic Lagrangian coherent structures~\citep{Haller:2015aa} of the flow. The complex intermingling of these hyperbolic LCS generates the stretching and folding motions that are characteristic of chaotic advection. The random nature of heterogeneous Darcy flow means that non-hyperbolic LCSs such as KAM islands cannot occur, hence these flows are ergodic and hyperbolic.

\subsection{Implications for Transport, Mixing and Reactions}\label{sec:diffusion}

The ubiquity of chaotic advection in heterogeneous Darcy flow has significant implications for the many fluidic processes hosted in this media, including transport and mixing of diffusive species such as solutes, colloids, reactive species and microorganisms. For these processes, the impact of chaotic dynamics scales with the P\'{e}clet number $Pe$ as $\sqrt{Pe}$~\citep{Aref:2017aa}, which can be large as $Pe$ ranges from $10^{-1}$ to $10^7$ in Darcy flows~\citep{Delgado:2007aa,Bear:1972aa}.

Chaotic advection leads to qualitative changes in solute mixing and transport. \citet{Le-Borgne:2013aa} use a lamellar mixing model~\citep{Duplat:2008aa} to show that in non-chaotic steady 2D Darcy flow, the rate of mixing of diffusive solutes is governed by the rate of fluid deformation imparted by the medium heterogeneity. For 2D~\citep{Dentz:2016aa} and 3D~\citep{Lester:2021aa} isotropic Darcy flow, the rate of elongation $\rho(t)\equiv l(t)/l(0)$ of of fluid elements grows algebraically as $ \langle \rho(t)\rangle \sim t^r$,
where the index $r=\mu+\nu$ varies from sublinear $r<1$ to ballistic stretching $1<r<2$, depending upon the medium heterogeneity, and $\mu$, $\nu$ respectively characterise the mean and variance of the fluid stretching process. The lamellar mixing model~\citep{Le-Borgne:2013aa} predicts that the concentration variance within the solute plume decays with time as
 \begin{align}
     &\langle c^2\rangle \propto t^{-2\mu+2\nu+D_f+2},
 \end{align}
where $D_f$ is the fractal dimension of material lines elongated by the flow. The lamellar mixing model also applies to turbulent and chaotic flows~\citep{Villermaux:2003ab,Meunier:2010aa}, yielding exponential decay of concentration variance as
 \begin{align}
     &\langle c^2\rangle \propto \left(\frac{\lambda_\infty}{6t}\right)^{1/4}e^{-\lambda_\infty t/3}.
 \end{align}
Hence mixing of diffusive solutes is rapidly accelerated by chaotic advection. Similarly, the peak concentration $c_m(t)$ of a Gaussian plume which is inversely proportional to the dilution index $E(t)\equiv\exp[H(t)]$~\citep{Kitanidis:1994aa} where $H(t)$ is the scalar entropy
\begin{equation}
H(t)=-\int_V d\mathbf{x} \frac{c(\mathbf{x},t)}{\langle c\rangle}\ln\left[\frac{c(\mathbf{x},t)}{\langle c\rangle}\right],
\end{equation}
decays algebraically in non-chaotic heterogeneous porous media flow as~\citep{Dentz:2018aa}
\begin{equation}
c_m(t)\propto\frac{1}{E(t)}\propto t^{-r},
\end{equation}
with $1<r<2$~\citep{Dentz:2016aa}, whereas for chaotic flows these quantities evolve exponentially as
\begin{equation}
c_m(t)\propto\frac{1}{E(t)}\propto \exp(-\lambda_\infty t).
\end{equation}
Hence chaotic advection accelerates the rate of solute mixing from algebraic to exponential.

Chaotic advection also significantly enhances transverse dispersion. In the purely advective limit $Pe\rightarrow\infty$, transverse dispersivity $D_{T,\infty}$ is zero in non-chaotic Darcy flow, whereas for chaotic flow $D_{T,\infty}\sim\lambda_\infty^2$ as per (\ref{eqn:model}). For diffusive solutes, $D_T$ in non-chaotic Darcy flow is proportional to molecular diffusivity $D_m$ as~\citep{Lester:2023aa}
\begin{equation}
D_T=D_m\langle m\rangle,\label{eqn:disp_isotropic}
\end{equation}
where $m$ characterises fluctuations in streamlines due to heterogeneities in the conductivity field. 
A simple model of stretching-mediated dispersion around streamlines in chaotic Darcy flow that yields conservative dispersion estimates (see Appendix~\ref{app:trans_disp_model} for details) yields significantly larger transverse dispersivity than (\ref{eqn:disp_isotropic})
\begin{equation}
D_T=D_{T,\infty}+D_m\frac{\sinh(2\lambda_\infty t_d)}{2\lambda_\infty t_d},\label{eqn:disp_anisotropic}
\end{equation}
where $t_d$ is the period of stretching events.
Conversely, chaotic mixing suppresses longitudinal dispersion. Although results are not available for heterogeneous Darcy flow, studies of the impact of chaotic mixing on hold-up dispersion~\citep{Jones:1994aa,Lester:2018ab} act as a guide for strongly heterogeneous media. For purely advective solutes, chaotic advection suppresses longitudinal variance from $\sigma^2_L(t)\sim t^2$~\citep{Taylor:1953aa} for non-chaotic flows to $\sigma^2_L(t)\sim t^2/(\ln t)^3$~\citep{Lester:2010aa} due to increased sampling of advective velocities. 

The impact of chaotic advection upon chemical reactions and biological activity is profound and multifaceted~\citep{Neufeld:2009aa}. For the simple case of fast binary reactions with Damkh\"{o}ler number $Da\rightarrow\infty$, the effective reaction kinetics are governed by solute mixing~\citep{Valocchi:2019aa,LeBorgne:2014aa}, hence the impacts outlined above for solute mixing directly translate to the rate and extent of these reactions. For simple binary reactions with finite $Da$, this acceleration becomes less pronounced with decreasing $Da$ as the advective dynamics play less of a controlling role. Chaotic advection also has a profound impact upon more complex chemical, biological and geological reaction systems hosted in porous materials such as autocatalytic reactions, oscillatory reactions, bistable and competitive systems. Although these reaction systems converge toward a stable or stationary chemical state under well-mixed conditions, under incomplete mixing conditions often found in porous materials~\citep{Wright:2017ab}, transport of reactants can continually drive this system away from its local equilibrium~\citep{Neufeld:2009aa}. Hence the accelerated transport characteristics inherent to chaotic advection can fundamentally alter the dynamics of these systems~\citep{Tel:2005aa}. Furthermore, the transport structures (LCSs) generated by these flows leading to qualitatively different macroscopic behaviour including, e.g., singularly enhanced reactions and altered stability of competitive species~\citep{Karolyi:2000aa}.

\section{Conclusions}\label{sec:conclusions}

The prevalence of chaotic advection in heterogeneous Darcy flow is a key consideration as these kinematics profoundly impact the myriad fluid-borne processes in porous media, ranging from solute mixing and transport to colloidal transport, chemical reactions and biological activity~\citep{Aref:2017aa}. In this study we directly address the question of the existence of chaotic advection for steady 3D Darcy flow with smooth, finite hydraulic conductivity fields and find that chaotic advection is ubiquitous for all realistic models of heterogeneous porous media.

We establish that realistic models of Darcy scale heterogeneous porous media should possess anisotropic hydraulic conductivity tensor fields $\mathbf{K}(\mathbf{x})$. We rely on two results. First, experimental observations~\cite{Bear:1972aa,Delgado:2007aa} show that transverse dispersion is ubiquitous in heterogeneous Darcy flow in the large P\'{e}clet number limit, i.e., purely advective conditions. Second, it has recently been proven~\cite{Lester:2023aa} that isotropic conductivity fields cannot admit purely advective transverse dispersion. Hence all realistic heterogeneous conductivity fields must also be anisotropic.

A recently uncovered quantitative relationship (\ref{eqn:model})~\citep{Lester:2024aa} between the purely advective transverse dispersivity $D_T$ and Lyapunov exponent $\lambda_\infty$ in random unidirectional 3 dof flows also extends to steady 3D Darcy flow. This establishes that transverse dispersion and chaotic advection in steady 3D heterogeneous Darcy flow are intimately linked as both phenomena result from non-trivial streamline braiding, hence chaotic advection is inherent to these flows. 

The onset of chaotic advection in steady 3D Darcy flow is considered via numerical simulations of flow in two classes of hydraulic conductivity fields; heterogeneous fields with variable anisotropy, and anisotropic fields with variable heterogeneity. We find that chaotic advection arises even for the weakest perturbations away from both isotropic heterogeneous media and anisotropic homogeneous media, establishing that chaotic advection is inherent to anisotropic heterogeneous media. Simple relationships are found for how $D_T$ and $\lambda_\infty$ scale with medium anisotropy parameter $\delta$ and conductivity log-variance $\sigma^2_{\ln K}$, and excellent agreement is found with the theoretical model (\ref{eqn:model}) linking $D_T$ and $\lambda_\infty$ in randomly braiding 3 dof flows. In the limit of large $\sigma^2_{\ln K}$, the Lyapunov exponent of anisotropic media converges to $\lambda_\infty\approx 0.6772$, which is close to the theoretical upper bound $\lambda_{\infty,\max}=\ln 2\approx 0.6931$ for 3 dof continuous systems, suggesting that highly heterogeneous Darcy flow is a strong mixing flow. 

These results firmly establish the ubiquity of chaotic advection in steady 3D heterogeneous Darcy flow, and we show that these kinematics have profound implications for understanding, quantifying and predicting a wide range of processes at the Darcy scale. 
The main finding of this study points to several future research directions. Further investigation of the quantitative link (\ref{eqn:model}) between $D_T$ and $\lambda_\infty$ is required, as $D_T$ has been measured for a wide range of porous materials~\citep{Delgado:2007aa}, whereas $\lambda_\infty$ has only been
measured in a small number of studies~\citep{Kree:2017aa,Souzy:2020aa,Heyman:2020aa,Heyman:2021aa} at the pore scale. Development of methods to characterize chaotic mixing at the field scale are required, as well as investigation of the chaotic dynamics generated by geologically-relevant conductivity fields, and fields that generate anomalous transport.
In the context of solute mixing and transport, further investigation and quantitative prediction of concentration PDF, transverse and longitudinal dispersion of diffusive solutes is required. Furthermore, the impact of chaotic mixing upon chemical reactions and biological activity in Darcy flow is an open area.

The recognition that chaotic dynamics are inherent to porous media flow across all scales opens the door to the development of a broad class of upscaling methods that explicitly honour these kinematics. The ubiquity of macroscopic chaotic advection has profound implications for the myriad processes hosted in heterogeneous porous media, and calls for a fundamental re-evaluation of transport and reaction methods in macroscopic porous systems.

\appendix

\section{Proof of Zero Total Helicity $H$}\label{app:zeroH}

We show that the total helicity $H$ for conductivity fields of the form
\begin{equation}
\begin{split}
    \mathbf{K}(\mathbf{x})=&\mathbf{K}_1(\mathbf{x})+\mathbf{K}_2(\mathbf{x}),\\
    =&k_0(\mathbf{x})\mathbf{I}+\delta k_0(\mathbf{x})\hat{\mathbf{e}}_1\otimes\hat{\mathbf{e}}_1,
\end{split}
\end{equation}
are zero by expressing the corresponding velocity field as
\begin{equation}
    \begin{split}
        \mathbf{v}(\mathbf{x})=&-\mathbf{K}_1(\mathbf{x})\cdot\nabla\phi-\mathbf{K}_2(\mathbf{x})\cdot\nabla\phi,\\
        =&\mathbf{v}_1(\mathbf{x})+\mathbf{v}_2(\mathbf{x}),
    \end{split}
\end{equation}
and thus the helicity density is
\begin{equation}
    \begin{split}
        \mathcal{H}(\mathbf{x})=&
        \mathbf{v}_1(\mathbf{x})\cdot[\nabla\times\mathbf{v}_1(\mathbf{x})]
        +\mathbf{v}_2(\mathbf{x})\cdot[\nabla\times\mathbf{v}_2(\mathbf{x})]\\
        +&\mathbf{v}_1(\mathbf{x})\cdot[\nabla\times\mathbf{v}_2(\mathbf{x})]+\mathbf{v}_2(\mathbf{x})\cdot[\nabla\times\mathbf{v}_1(\mathbf{x})],\label{eqn:epi_helicity}
    \end{split}
\end{equation}
where the quantities on the top row of the RHS are zero. Expressing $\mathbf{v}_2(\mathbf{x})=f(\mathbf{x})\hat{\mathbf{e}}_1$, then (\ref{eqn:epi_helicity}) simplifies to
\begin{equation}
\begin{split}
    \mathcal{H}(\mathbf{x})=&f(\mathbf{x})\hat{\mathbf{e}}_1\cdot[\nabla\times\mathbf{v}_1(\mathbf{x})]+\mathbf{v}_1(\mathbf{x})\cdot[\nabla\times(f(\mathbf{x})\hat{\mathbf{e}}_1)]\\
    =&\hat{\mathbf{e}}_1\cdot(\nabla\times[f(\mathbf{x})\mathbf{v}_1(\mathbf{x})])\\
    =&\nabla\cdot(x_1\nabla\times[f(\mathbf{x})\mathbf{v}_1(\mathbf{x})]).
\end{split}
\end{equation}
Hence the total helicity
\begin{equation}
    H=\int_\Omega \mathcal{H}(\mathbf{x})d\mathbf{x}=0,
\end{equation}
due to the divergence theorem and continuity of $f(\mathbf{x})$ and $\mathbf{v}_1(\mathbf{x})$ over the boundary $\partial\Omega$.

\section{Fluid Deformation in Protean Coordinates}\label{app:Lyapunov}

Fluid deformation is characterised by the deformation gradient tensor $\mathbf{F}(t;\mathbf{X})$ with Lagrangian coordinate $\mathbf{X}$, which evolves according to (\ref{eqn:deform}).
The Lyapunov exponent and fluid stretching statistics are gathered by sampling $\boldsymbol\epsilon$ along streamlines. The moving and rotating \emph{Protean} coordinate frame provides several advantages as detailed in \citep{Lester:2018aa} and briefly summarised as follows. The Protean coordinate frame $\mathbf{x}^\prime$ is related to the Eulerian frame $\mathbf{x}$ as
\begin{equation}
\mathbf{x}^\prime(t)=\mathbf{Q}(t)\cdot[\mathbf{x}-\mathbf{x}_0(t;\mathbf{X})],
\end{equation}
 where $\mathbf{x}_0(t;\mathbf{X})$ is the position of a fluid tracer particle at time $t$ with initial position $\mathbf{X}$, and $\mathbf{Q}(t)$ is a time-dependent orthogonal rotation matrix. The Protean velocity gradient tensor $\boldsymbol\epsilon^\prime(t;\mathbf{X})$ is related to the Lagrangian velocity gradient tensor $\boldsymbol\epsilon(t)$ as
\begin{equation}
\boldsymbol\epsilon^\prime(t;\mathbf{X})=\mathbf{Q}^\top(t)\cdot\boldsymbol\epsilon(t;\mathbf{X})\cdot\mathbf{Q}(t)+\dot{\mathbf{Q}}^\top(t)\cdot\mathbf{Q}(t),\label{eqn:Protean_eps}
\end{equation}
The rotational matrix $\mathbf{Q}(t)$ aligns the $x_1^\prime$ coordinate with the local velocity direction such that $\hat{\mathbf{e}}_1^\prime=\mathbf{v}/v$, and for steady flow the Protean coordinate system is a streamline coordinate system. The rotation $\mathbf{Q}(t)$ is comprised of two subrotations as $\mathbf{Q}(t)=\mathbf{Q}_2(t)\cdot\mathbf{Q}_1(t)$, where the first rotation $\mathbf{Q}_1(t)$ aligns $x_1^\prime$ with $\mathbf{v}/v$ and so renders the $\epsilon_{21}^\prime$ and $\epsilon_{31}^\prime$ elements of the Protean velocity gradient tensor $\boldsymbol\epsilon^\prime(t;\mathbf{X})$ to be zero~\citep{Lester:2018aa}. The second rotation $\mathbf{Q}_2(t)$ about the axis $\mathbf{x}^\prime$ in the streamwise direction is chosen such the remaining lower triangular element $\epsilon_{23}^\prime$ is also zero~\citep{Lester:2018aa}, rendering the Protean velocity gradient tensor upper triangular:
\begin{equation}
   \boldsymbol\epsilon^\prime=\left(
\begin{array}{ccc}
     \epsilon^\prime_{11} & \epsilon^\prime_{12} & \epsilon^\prime_{13} \\
    0 & \epsilon^\prime_{22} & \epsilon^\prime_{23}\\
    0 & 0 & \epsilon^\prime_{33}
\end{array}
    \right).
\end{equation}
Hence fluid deformation in the heterogeneous Darcy flows is characterised by computation of the Protean velocity gradient tensor along the streamlines of these flows.


\section{Fluid Stretching in 3D Darcy Flow}\label{app:1Dstretch}

The 1D CTRW (\ref{eqn:ctrw}) for evolution of $\rho(t;\mathbf{X})$ is derived by considering evolution of the
infinitesimal line element
$\delta \mathbf{l}(t;\mathbf{X})=\mathbf{F}^\prime(t;\mathbf{X})\cdot\delta \mathbf{l}(0;\mathbf{X})$, the length of which evolves as
\begin{equation}
\delta l(t;\mathbf{X})\equiv| |\delta \mathbf{l}(t;\mathbf{X})|=\sqrt{ \mathbf{l}(0;\mathbf{X})\cdot\mathbf{C}(t;\mathbf{X})\cdot \mathbf{l}(0;\mathbf{X})},
\end{equation}
where $\mathbf{C}(t;\mathbf{X})=\mathbf{F}(t;\mathbf{X})^\top\cdot \mathbf{F}(t;\mathbf{X})$ is the symmetric Cauchy-Green tensor. Hence $\rho(t;\mathbf{X})$ grows with the largest eigenvalue $\nu(t;\mathbf{X})$ of $\mathbf{C}(t;\mathbf{X})$ as 
\begin{equation}
\rho(t;\mathbf{X})=\sqrt{\nu(t,;\mathbf{X})}.
\end{equation}
As detailed in Appendix~\ref{app:Lyapunov}, the $F_{22}^\prime$ component characterises exponential stretching of fluid elements, and so converges to $\rho$~\citep{Lester:2018aa} in the asymptotic limit as
\begin{equation}
\lim_{t\rightarrow\infty}\frac{1}{t}\ln\rho(t;\mathbf{X})
=\lim_{t\rightarrow\infty}\frac{1}{2t}\ln\nu(t;\mathbf{X})
=\lim_{t\rightarrow\infty}\frac{1}{t}\ln F_{22}^\prime(t;\mathbf{X}).
\end{equation}
From (\ref{eqn:deform}), $F_{22}^\prime$ evolves with streamline distance $s$ as
\begin{equation}
F_{22,s}^\prime(s;\mathbf{X})=\exp\left(\int_0^s \frac{\epsilon^\prime_{22}(s^\prime;\mathbf{X})}{v(s^\prime;\mathbf{X})} ds^\prime\right),
\end{equation}
where $v(s;\mathbf{X})=|\mathbf{v}(\mathbf{x}(s;\mathbf{X}))|=ds/dt$ is the local streamline velocity and $\mathbf{x}(s;\mathbf{X})$ is the position of a tracer particle along a streamline with initial position $\mathbf{X}$ at $s=0$, $t=0$. As fluid velocity and velocity gradient are both spatially Markovian along streamlines~\citep{Le-Borgne:2008aa,Lester:2022aa}, $s$ may be discretized with respect to the spatial velocity correlation length $\ell$ as $s_n=n \ell$, with $n=0,1,\dots$, as per (\ref{eqn:ctrw}). Similarly, the advection time $t_n$ and length stretch $\rho_n\equiv F_{22,s}^\prime(s_n;\mathbf{X})$ at position $s_n$ along a streamline evolves as per (\ref{eqn:ctrw}), with $v_n\equiv v(s_n;\mathbf{X})$, $\epsilon_n\equiv\epsilon^\prime_{22}(s_n;\mathbf{X})$. Thus the CTRW (\ref{eqn:ctrw}) captures the line stretching dynamics in steady heterogeneous 3D Darcy flow.

\subsection{Fluid Stretching Under Normal Transport}\label{app:topo_gauss}

Following~\citep{Dentz:2016aa}, from the CTRW (\ref{eqn:ctrw}), the Fourier-Laplace transform $\hat{p}_{\zeta,t}(k,\lambda)$ of the PDF $p_{\zeta,t}(\zeta,t)$ where $\lambda$ is the Laplace variable for $\zeta$ and $k$ is the Fourier variable for $t$ is given by
 \begin{align}
\hat{p}_{\zeta,t}(k,\lambda) = \frac{1 - \psi(\lambda)}{\lambda} \frac{1}{1 - \psi(k,\lambda)},
 \end{align}
 where $\psi(k,\lambda)$ is given by
 \begin{align}
 \label{eq:psik}
\psi(k,\lambda) = \psi(\lambda) + \int\limits_0^\infty dt \exp(-\lambda t) [\cos(|k|\epsilon t) -1] \psi(t),  
 \end{align}
and $\psi(\lambda)$ is the Laplace transform of $\psi(\tau)$. The moments of $\zeta$ are defined in terms of $\hat{p}_{\zeta,t}(k,\lambda)$ by
 \begin{align}
m_{j,\zeta} = (-i)^j \frac{\partial^j \hat{p}_{\zeta,t}(k,\lambda)}{\partial k^j}\Big|_{k = 0}. 
 \end{align}
 Hence
 \begin{align}
m_{1,\zeta} = 0, && m_{2,\zeta} = -\frac{1}{\lambda [1 - \psi(\lambda)]}\frac{\partial^2 \psi(k,\lambda)}{\partial k^2}\Big|_{k = 0},
 \end{align}
and the small $k$-expansion of~\eqref{eq:psik} is  
 \begin{align}
 \label{eq:psik2}
\psi(k,\lambda) = \psi(\lambda) - \int\limits_0^\infty dt \exp(-\lambda t) \frac{1}{2} k^2 \sigma_\epsilon^2 t^2 \psi(t) + \dots 
 \end{align}
and additional small $\lambda$-expansion
 \begin{align}
 \label{eq:psik3}
\psi(k,\lambda) =  1 -  k^2 \sigma_\epsilon^2 \langle \tau^2 \rangle + \dots  
 \end{align}
The small $\lambda$-expansion of $\psi(\lambda)$ is 
 \begin{align}
 \label{eq:psi}
\psi(\lambda) =  1 - \langle \tau \rangle \lambda + \frac{1}{2} \langle \tau^2 \rangle \lambda^2,
 \end{align}
thus, we obtain to leading order
 \begin{align}
m_{2,\zeta} = \frac{\sigma_\epsilon^2\langle \tau^2 \rangle}{\lambda^2\langle \tau \rangle},
 \end{align}
the inverse Laplace transform of which gives
 \begin{align}
\lim_{t\rightarrow\infty}\sigma^2_{\ln\rho}  = \sigma_\epsilon^2\frac{\langle \tau^2 \rangle}{\langle \tau \rangle}t\equiv\sigma^2_\lambda t. 
 \end{align}

\section{Numerical Solvers and Streamline Tracking}\label{app:numerics}

The potential fluctuation equation (\ref{eqn:potfluc}) for Darcy flow is solved to precision $10^{-16}$ on a uniform structured $256^3$ grid using a high resolution eighth-order compact finite difference scheme~\citep{LeLe:1992aa}. To generate high resolution results and preserve the Lagrangian kinematics of the flow, we use a similar numerical approach to that used in \citep{Lester:2019ab}. Specifically, we perform a triply-periodic 5-th order spline interpolation of the primitive variables $\tilde{\phi}(\mathbf{x})$, $\mathbf{K}(\mathbf{x})$ from their grid values and reconstruct the potential field $\phi(\mathbf{x})$ according to (\ref{eqn:phidecomp}). The velocity field is then computed analytically from these fields as $\mathbf{v}(\mathbf{x})=-\mathbf{K}(\mathbf{x})\cdot\nabla\phi(\mathbf{x})$, ensuring that the velocity field is triply-periodic and $C_4$ continuous and the velocity gradient is accurately resolved for computation of fluid deformation in the Protean frame. This approach also implicitly enforces the helicity-free constraint $h(\mathbf{x})=0$ for the case of isotropic conductivity tensor $\mathbf{K}(\mathbf{x})=k(\mathbf{x})\mathbf{I}$ for $\delta=0$. For the $256^3$ mesh the local relative divergence error 
\begin{equation}
    d(\mathbf{x})=\frac{|\nabla\cdot\mathbf{v}(\mathbf{x})|}{||\nabla\mathbf{v}(\mathbf{x})||},
\end{equation}
of the interpolated velocity field $\mathbf{v}(\mathbf{x})$ is order $10^{-4}$ and the velocity gradient is accurate to order $10^{-3}$. 

Streamline tracking is then computed by solving the advection equation from the initial Lagrangian coordinate $\mathbf{X}$ as
\begin{equation}
    \frac{d\mathbf{x}}{dt}=\mathbf{v}(\mathbf{x}(t;\mathbf{X})),\quad\mathbf{x}(0;\mathbf{X})=\mathbf{X},\label{eqn:advection}
\end{equation}
via a 5-th order Cash-Karp Runge-Kutta scheme to precision $10^{-14}$. The periodic boundaries allow advection of fluid streamlines over many multiples of the solution domain $\Omega$, facilitating study of the Lagrangian kinematics over arbitrary distances. Although the corresponding velocity field is periodic in space, when the flow is chaotic the streamlines are aperiodic and eventually sample all of the conductivity field in an ergodic manner.\\

While accurate, this streamline tracking method (along with all numerical schemes which do not explicitly enforce kinematic constraints) has been shown~\citep{Lester:2023aa} to introduce spurious transverse dispersion for the isotropic zero helicity density flow $h=0$ due to numerical streamlines not follwing their analytic counterparts. To circumvent this problem for the helicity-free case $\delta=0$, we instead solve the invariant streamfunctions $\psi_1(\mathbf{x})$, $\psi_2(\mathbf{x})$ for the velocity field $\mathbf{v}(\mathbf{x})=\nabla\psi_1(\mathbf{x})\times\nabla\psi_2(\mathbf{x})$ via the following governing equations~\citep{Lester:2022aa} to precision $10^{-16}$ using the same finite-difference method as described above:
\begin{align}
\nabla^2\psi_1(\mathbf{x})-\nabla f(\mathbf{x})\cdot\nabla\psi_1(\mathbf{x})=S_1(\psi_1,\psi_2),\\
\nabla^2\psi_2(\mathbf{x})-\nabla f(\mathbf{x})\cdot\nabla\psi_2(\mathbf{x})=S_2(\psi_1,\psi_2),
\label{eqn:sfuncs}
\end{align}
where $f=\ln k$ and 
\begin{align}
S_1=\frac{(\mathbf{B}\times\psi_2)\cdot(\nabla\psi_1\times\nabla\psi_2)}{|\nabla\psi_1\times\nabla\psi_2|} &&
S_2=\frac{(\mathbf{B}\times\psi_1)\cdot(\nabla\psi_1\times\nabla\psi_2)}{|\nabla\psi_1\times\nabla\psi_2|},
\end{align}
and
\begin{equation}
    \mathbf{B}\equiv(\nabla\psi_1\cdot\nabla)\nabla\psi_2)-(\nabla\psi_2\cdot\nabla)\nabla\psi_1).
\end{equation}
Similar to the Darcy equation, continuous streamfunctions $\psi_1(\mathbf{x})$, $\psi_2(\mathbf{x})$ are reconstructed from grid data using triply-periodic splines and the velocity field is computed analytically from these streamfunctions. As shown in \citep{Lester:2022aa}, this method yields the same velocity field (to within numerical error) as that given by direct solution of the Darcy equation.\\

Each family of streamfunctions $\psi_i$ is comprised of a \emph{foliation} of non-intersecting streamsurfaces $\psi_i=\text{const.}$ that span the flow domain and constrain the Lagrangian kinematics of the flow. This flow structure is non-chaotic as the advection equation (\ref{eqn:advection}) simplifies to
\begin{equation}
\frac{ds}{dt}=v(s;\psi_1(\mathbf{X}),\psi_2(\mathbf{X}),\quad s(t=0;\mathbf{X})=0,
\label{eqn:advect}
\end{equation}
where $s$ is the distance travelled along a streamline of a tracer particle with initial position $\mathbf{X}$. The velocity magnitude $v(s;\psi_1(\mathbf{X}),\psi_2(\mathbf{X})$ at the intersection of the streamsurfaces $\psi_1(\mathbf{x})=\psi_1(\mathbf{X})$, $\psi_2(\mathbf{x})=\psi_2(\mathbf{X})$ only varies with $s$. Equation (\ref{eqn:advect}) is \emph{integrable} in that $\psi_1$, $\psi_2$ represent two invariants of the flow in the 3D domain, resulting in only one degree of freedom (distance) for streamlines of the flow to explore. For helicity-free flow, we perform streamline tracking via numerical integration of (\ref{eqn:advect}) to precision $10^{-8}$. This approach ensures numerical streamlines follow their analytic counterparts and so enforces zero transverse dispersion and prevents the non-trivial braiding of streamlines that lead to chaotic advection.

\begin{figure}
\begin{centering}
\begin{tabular}{c c}
\includegraphics[width=0.45\columnwidth]{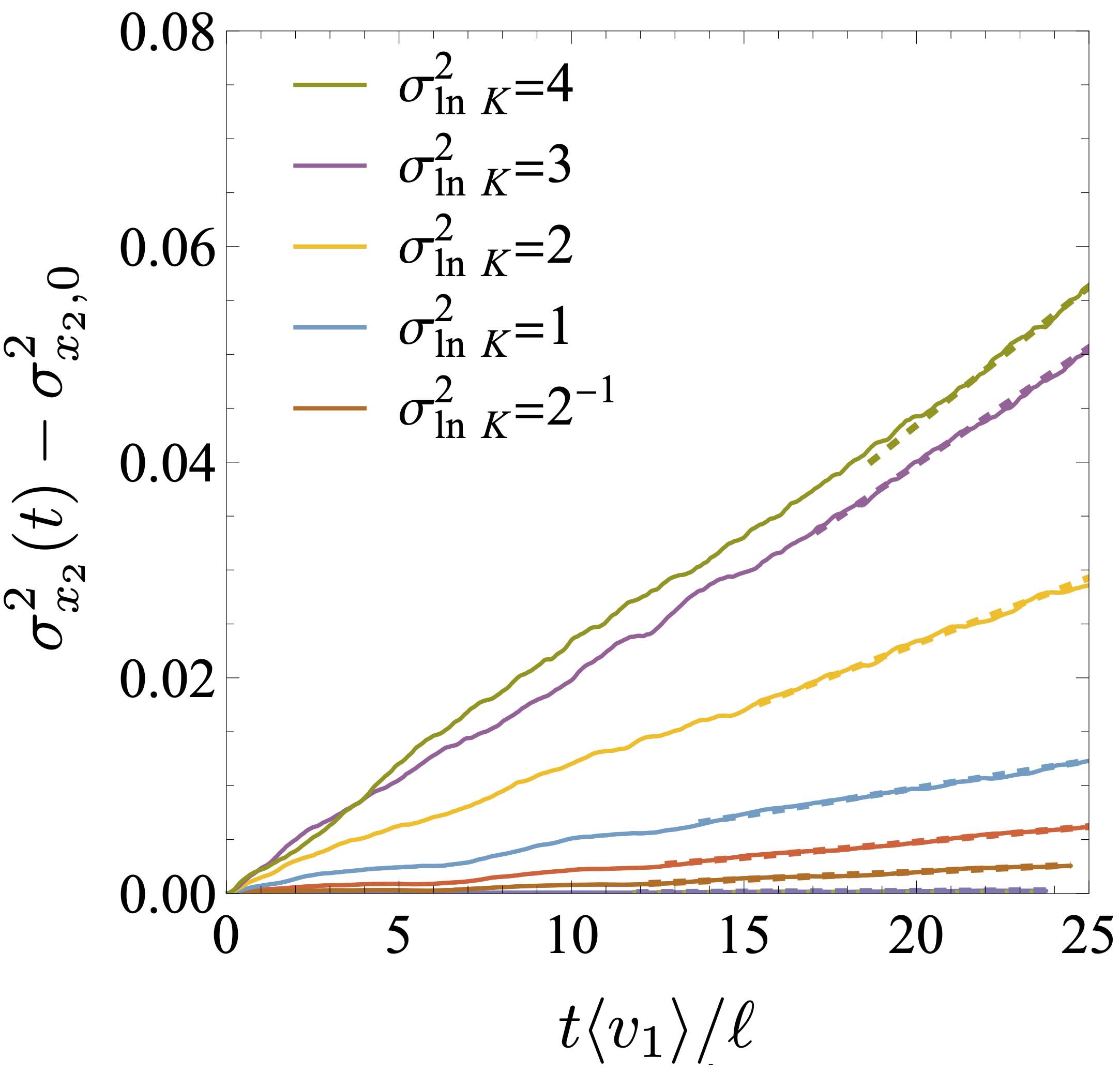}&
\includegraphics[width=0.45\columnwidth]{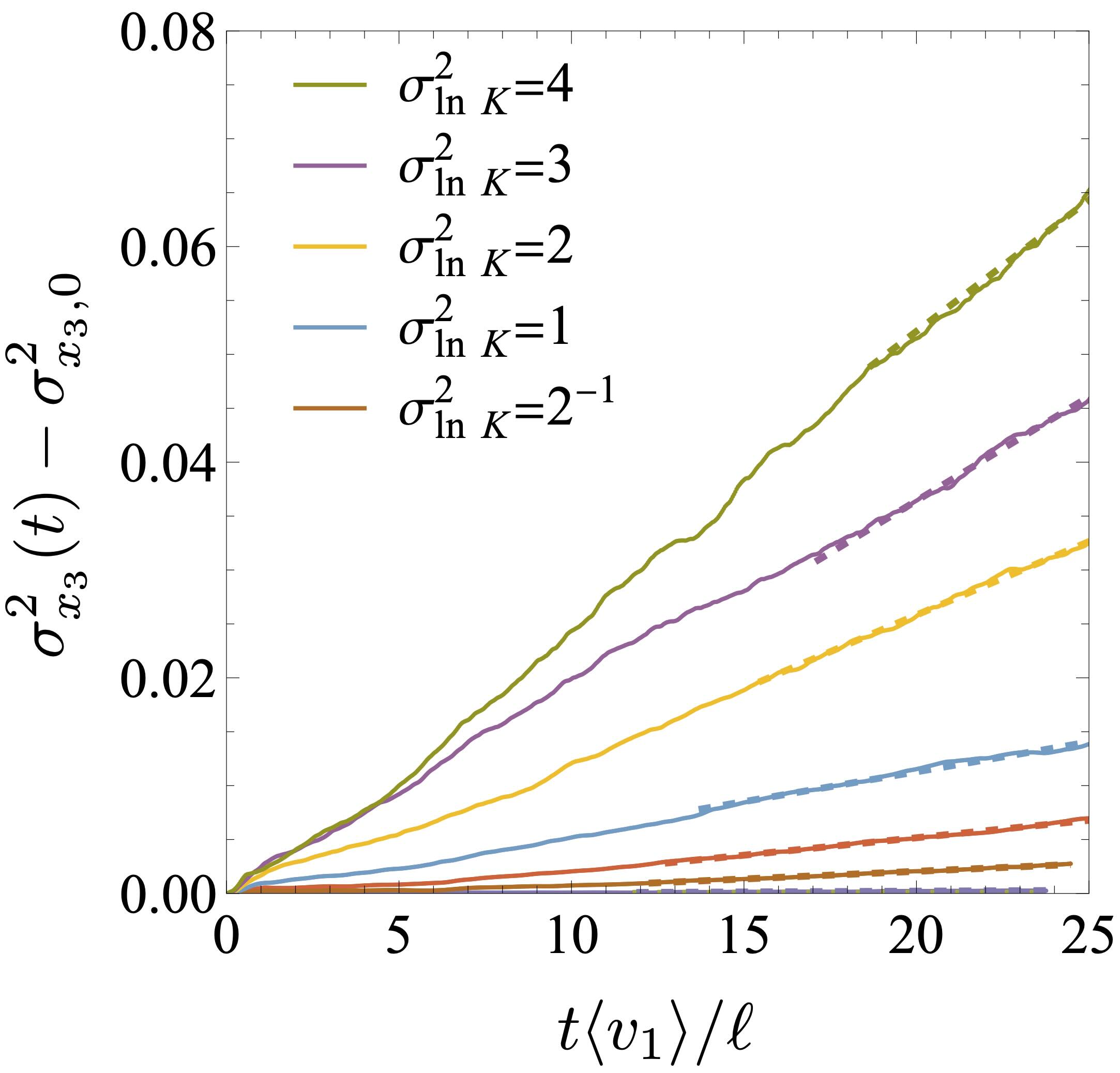}\\
(a) & (b)\\
\includegraphics[width=0.48\columnwidth]{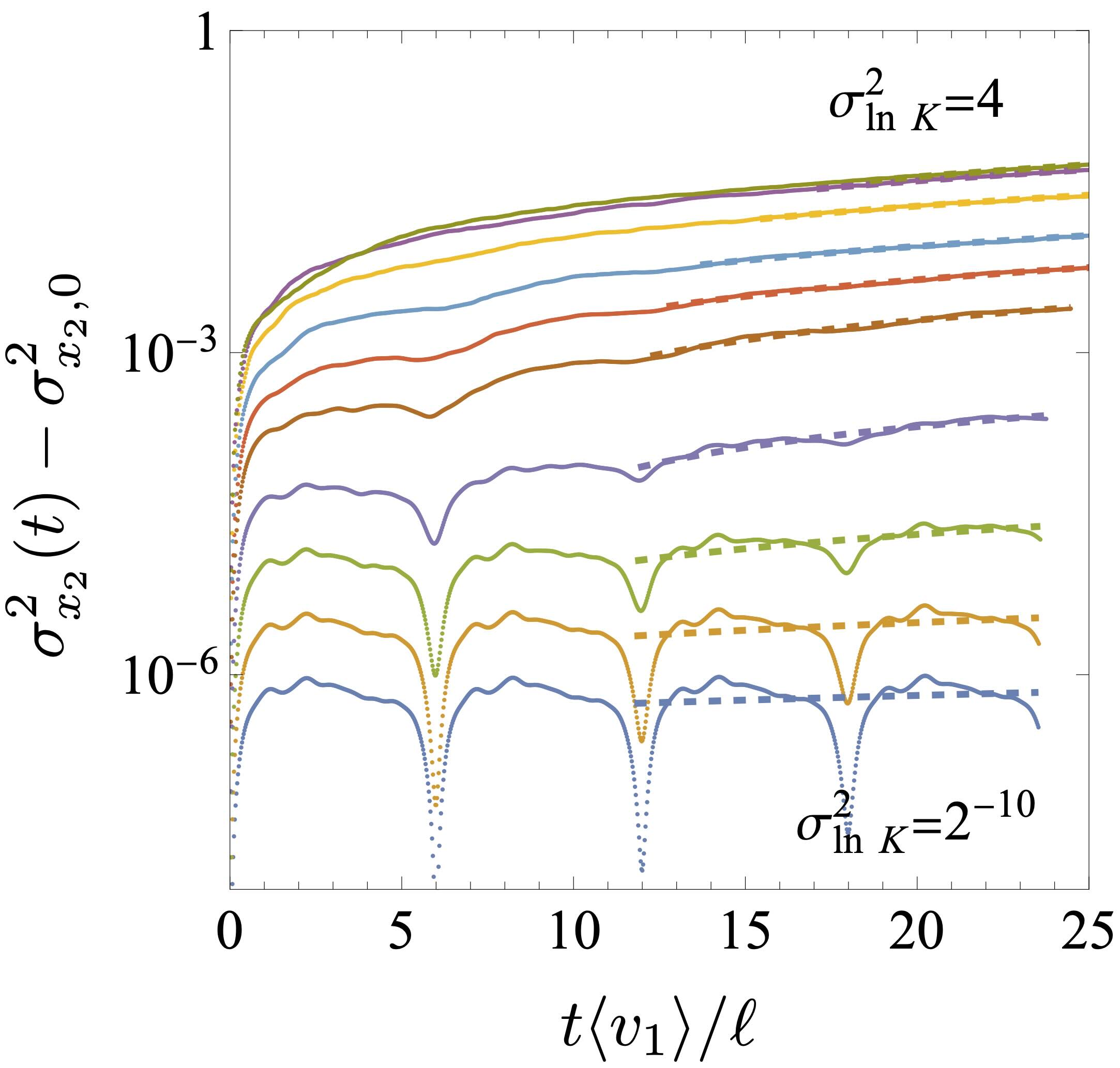}&
\includegraphics[width=0.48\columnwidth]{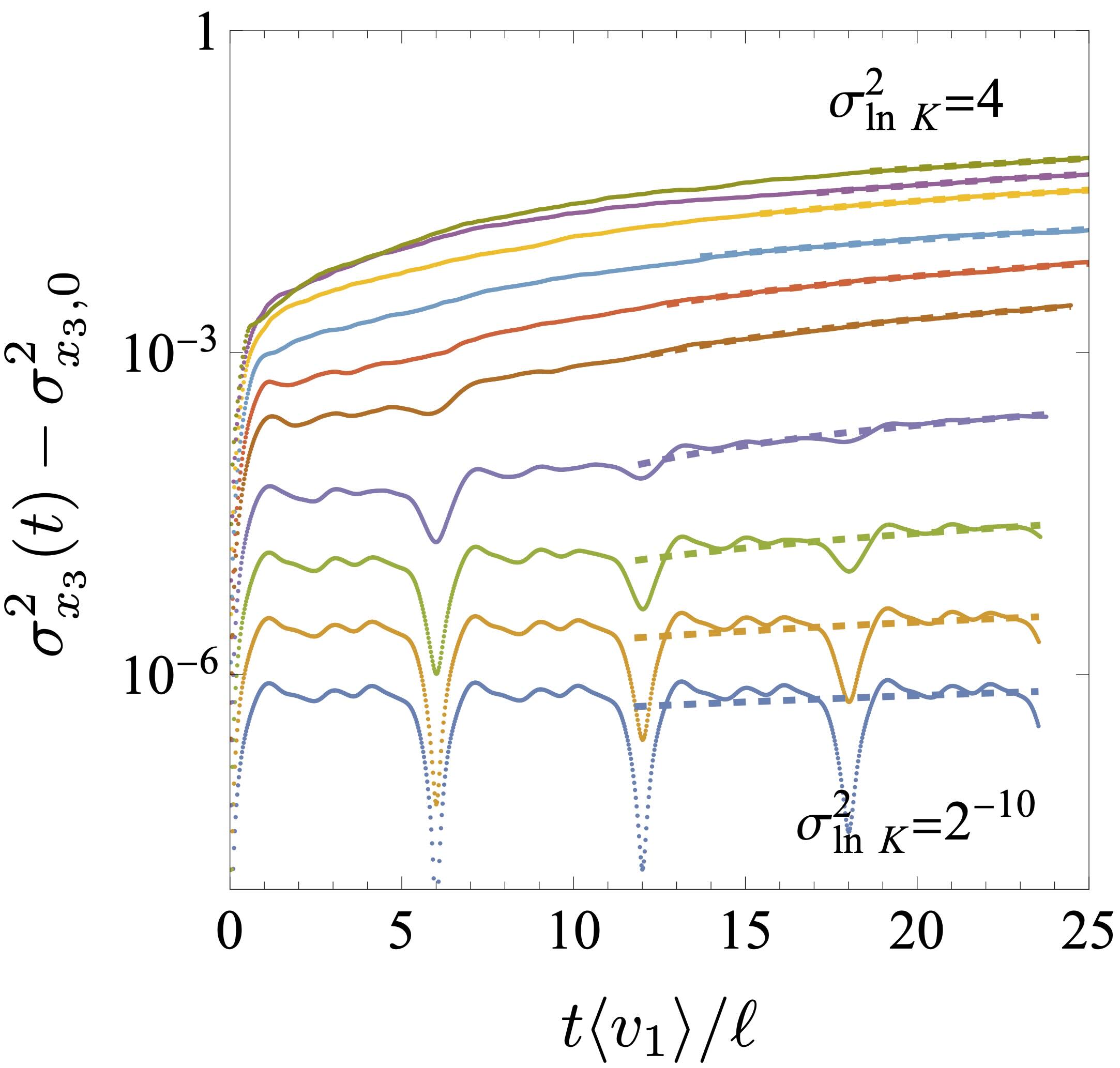}\\
(c) & (d)
\end{tabular}
\end{centering}
\caption{Evolution of transverse scalar variances $\sigma_{x_2}^2(t)$ (left), $\sigma_{x_3}^2(t)$ (right) with dimensionless travel time $t\langle v_1\rangle/\ell$ in linear (top) and logarithmic (bottom) scales for heterogeneous anisotropic Darcy flow for various values of log-variance $\sigma_{\ln K}^2$. Fitted linear trend (dashed lines) at late times is used to estimate transverse dispersivities $D_{22}$, $D_{33}$.}
\label{fig:variances}
\end{figure}

\section{Calculation of Transverse Dispersivity}\label{app:dispersivity}

For both the variable anisotropy and variable heterogeneity Darcy flows, transverse dispersivity is determined by tracking $N_p=10^3$ streamlines over $10^3$ traverses of the periodic domain $\Omega$ seeded from random locations within $\Omega$. From these streamlines, the transverse variances are computed as
\begin{align}&\sigma_{x_2}^2(t)=\frac{1}{N_p}\sum_{i=1}^{N_p}(x_{2,i}(t)-\langle x_2\rangle(t))^2,\quad \langle x_2\rangle(t)=\frac{1}{N_p}\sum_{i=1}^{N_p}x_{2,i}(t),\\&\sigma_{x_3}^2(t)=\frac{1}{N_p}\sum_{i=1}^{N_p}(x_{3,i}(t)-\langle x_3\rangle(t))^2,\quad \langle x_2\rangle(t)=\frac{1}{N_p}\sum_{i=1}^{N_p}x_{2,i}(t).
\end{align}
For the anisotropic Darcy flow with variable heterogeneity, the transverse variances exhibit slow growth and periodic oscillations in Figure~\ref{fig:variances} for weak conductivity variance $\sigma^2_{\ln K}\ll 1$ as the streamlines of the flow are nearly periodic. With increasing heterogeneity $\sigma^2_{\ln K}$, these orbits lose periodicity and ergodically explore the flow domain, resulting in stronger growth of the transverse variances. The transverse dispersivities are related to the asymptotic variance growth as
\begin{align}
    D_{22}=\frac{1}{2}\lim_{t\rightarrow\infty}\frac{d\sigma_{x_2}^2}{dt} &&
    D_{33}=\frac{1}{2}\lim_{t\rightarrow\infty}\frac{d\sigma_{x_3}^2}{dt}. 
\end{align}
To estimate this asymptotic growth we fit a linear function to the asymptotic variance data over two periods of the flow, as shown in Figure~\ref{fig:variances}. Note that the variance data in Figure~\ref{fig:variances} has been truncated to 20 correlation lengths for illustrative purposes. The total transverse dispersivity $D_T$ is then computed as the average of the $x_2$ $x_3$ dispersivities.



\section{Estimate of Diffusive Transverse Dispersivity}\label{app:trans_disp_model}

To generate a conservative estimate of diffusive transverse dispersivity $D_T$ under the action of chaotic advection, we consider a simple model for fluid deformation around at streamline at Lagrangian position $\mathbf{X}_0$ oriented in the $\mathbf{v}/v=\hat{\mathbf{e}}^\prime_1$ direction of the Protean frame. Evolution of the local fluid deformation gradient tensor along the streamline is modeled via a ``stretch and relax'' process over period $t_d$ as
\begin{equation}
    \mathbf{F}_0(t)=\begin{cases}
        \mathbf{F}_{00}(t)\quad\text{for}\,\, &0<t\leqslant t_d,\\
        \mathbf{F}_{00}(t_d-t)\quad\text{for}\,\, &t_d<t\leqslant 2t_d,\\
    \end{cases}
\end{equation}
and $\mathbf{F}_{00}(t)=
\exp\left(\lambda_\infty t\right)\hat{\mathbf{e}}^\prime_2\otimes\hat{\mathbf{e}}^\prime_2\,
+
\exp\left(-\lambda_\infty t\right)\,\hat{\mathbf{e}}^\prime_3\otimes\hat{\mathbf{e}}^\prime_3$ characterizes the transverse stretching and contraction process. The deformation gradient tensor $\mathbf{F}(\mathbf{X}_0,t)$ on a streamline with Lagrangian coordinate $\mathbf{X}_0$ is then modelled as a series of sequential ``stretch and relax'' processes with uniformly distributed random orientation $\theta\in[0,2\pi]$ as
\begin{equation}
    \mathbf{F}(\mathbf{X},t)=\left[\mathbf{R}(\theta_n)\cdot \mathbf{F}_0(t)\cdot\mathbf{R}(\theta_n)^\top\right]\cdot\,\dots\,\cdot\left[\mathbf{R}(\theta_1)\cdot \mathbf{F}_0(2t_d)\cdot\mathbf{R}(\theta_1)^\top\right]),\label{eqn:seqence}
\end{equation}
where $\mathbf{R}(\theta)$ is the rotation matrix around $\hat{\mathbf{e}}^\prime_1$ and $n=\lceil t/(2t_d)\rceil$. This simple model generates a conservative estimate of dispersion as it does not capture persistent exponential stretching of fluid elements.
However, such stretching in the absence of folding leads to erroneous predictions of transverse dispersion that grow exponentially in time. In practice, folding suppresses transverse dispersion to be sub-exponential, however incorporation of such second-order deformation measures is beyond the scope of this model.  

A solute packet centered about a streamline with Eulerian coordinate $\mathbf{x}_0(t)\equiv\mathbf{x}(\mathbf{X}_0,t)$ with initial concentration $c(\mathbf{x},0)=\delta(\mathbf{x}-\mathbf{x}_0)$ evolves with a Gaussian concentration profile
\begin{equation}
    c(\mathbf{x},t)=\frac{1}{\sqrt{(2\pi)^d\det\boldsymbol\Sigma(t)}}\exp\left(-\frac{1}{2}(\mathbf{x}-\mathbf{x}_0(t))\cdot\boldsymbol\Sigma^{-1}(t)\cdot(\mathbf{x}-\mathbf{x}_0(t))\right),\label{eqn:soln}
\end{equation}
where the covariance matrix $\boldsymbol\Sigma(t)$ evolves as~\citep{Dentz:2018aa}
\begin{align}
&\boldsymbol\Sigma(t)=2D_m\mathbf{F}^\prime(\mathbf{X}_0,t)\cdot\left(\int_0^t \mathbf{F}^\prime(\mathbf{X}_0,t^\prime)^{-1}\cdot\mathbf{F}^\prime(\mathbf{X}_0,t^\prime)^{-\top}dt^\prime\right)\cdot\mathbf{F}^\prime(\mathbf{X}_0,t)^\top,
\end{align}
which from (\ref{eqn:seqence}) simplifies to
\begin{equation}
    \boldsymbol\Sigma(2nt_d)=2D_m\sum_{i=1}^n\mathbf{R}(\theta_n)\cdot\boldsymbol\Lambda_0\cdot\mathbf{R}(\theta_n)^\top,
\end{equation}
where
\begin{equation}
    \boldsymbol\Lambda_0=
    \frac{1-e^{-2\hat{\lambda}_\infty t_d}}{\hat\lambda_\infty}\hat{\mathbf{e}}^\prime_2\otimes\hat{\mathbf{e}}^\prime_2+
\frac{e^{2\hat{\lambda}_\infty t_d}-1}{\hat\lambda_\infty}\hat{\mathbf{e}}^\prime_2\otimes\hat{\mathbf{e}}^\prime_2.
\end{equation}
Hence for large $t$, the covariance matrix converges to $\boldsymbol\Sigma(t)=2D_m n \langle\boldsymbol\Lambda\rangle$ where
\begin{equation}
    \langle\boldsymbol\Lambda\rangle\equiv\frac{1}{2\pi}\int_0^{2\pi}\left[\mathbf{R}(\theta)\cdot\boldsymbol\Lambda_0\cdot\mathbf{R}(\theta)^\top\right]d\theta
\end{equation}
and so
\begin{equation}
    \boldsymbol\Sigma(t)=2D_m t\frac{\sinh 2\hat{\lambda}_\infty t_d}{2\hat{\lambda}_\infty t_d}\mathbf{1}.
\end{equation}
Hence transverse dispersion around streamlines is exponentially amplified by periodic stretching. In conjunction with advective dispersion of streamlines, this yields (\ref{eqn:disp_anisotropic}). Note that similar to (\ref{eqn:disp_isotropic}), this model is only valid for $||\boldsymbol\Sigma(t)||<\ell$.

\bibliography{reflist}
\bibliographystyle{jfm}

\end{document}